\documentclass[aps,prd,amsmath,amssymb,preprintnumbers,nofootinbib,a4paper,11pt]{revtex4}
\pdfoutput=1
\usepackage{amsthm}
\usepackage{graphicx}
	\graphicspath{{./NewFig/}}
\usepackage{url}
\usepackage[bookmarks, pagebackref=false]{hyperref}
\usepackage{color}
	\definecolor{rossoCP3}{cmyk}{0,.88,.77,.40}
		\definecolor{graa}{rgb}{0.8,0.8,0.8}
		\definecolor{blaa}{rgb}{0.2,0.2,0.6}
		\hypersetup{
			colorlinks, 
			bookmarksopen, 
			bookmarksnumbered,
			citecolor=blaa, 		
			linkcolor=rossoCP3,	
			urlcolor=rossoCP3,			
			}
\usepackage{dcolumn}
\usepackage{bm}
\usepackage{bbm}
\usepackage{pxfonts}

\usepackage{epsfig}
\usepackage{placeins}

\usepackage{bbold}

\usepackage[ margin=5pt, font=small,labelfont=bf, justification=raggedright]{caption}
\usepackage{youngtab}
\usepackage{slashed}
\usepackage{subfig}

\newcommand{\beq}{\begin{eqnarray}}
\newcommand{\eeq}{\end{eqnarray}}

\newcommand{\bmp}{\noindent\begin{minipage}{16cm}}
\newcommand{\emp}{\end{minipage}\vskip 7mm} 

\def\lsim{\mathrel{\rlap{\lower4pt\hbox{\hskip1pt$\sim$}}
    \raise1pt\hbox{$<$}}}                
\def\gsim{\mathrel{\rlap{\lower4pt\hbox{\hskip1pt$\sim$}}
    \raise1pt\hbox{$>$}}}                

\begin{document}

\title{\LARGE \color{rossoCP3} Conformal Behavior at Four Loops and Scheme (In)Dependence}
 \author{Thomas A. Ryttov}\email{ryttov@cp3.dias.sdu.dk} 
  \affiliation{
{ \color{rossoCP3}  \rm CP}$^{\color{rossoCP3} \bf 3}${\color{rossoCP3}\rm-Origins} \& the Danish Institute for Advanced Study {\color{rossoCP3} \rm DIAS},\\ 
University of Southern Denmark, Campusvej 55, DK-5230 Odense M, Denmark.
}

\begin{abstract}
We search for infrared zeros of the beta function and evaluate the anomalous dimension of the mass at the associated fixed point for asymptotically free vector-like fermionic gauge theories with gauge group $SU(N)$. The fixed points of the beta function are studied at the two, three and four loop level in two different explicit  schemes. These are the modified regularization invariant, RI', scheme and the minimal momentum subtraction, mMOM, scheme. The search is performed in Landau gauge where the beta function of the gauge parameter vanishes. We then compare our findings to earlier identical investigations performed in the modified minimal subtraction, $\overline{\text{MS}}$, scheme. 

It is found that the value of the anomalous dimension of the mass is smaller at three and four loops than at two loops. This seems to be a generic pattern that is observed in all three different schemes. We then estimate the value of the anomalous dimension to be $\gamma \sim 0.225-0.375$ for twelve fundamental flavors and three colors, $\gamma \sim 0.500 - 0.593$ for two adjoint flavors and two colors and finally $\gamma \sim 1.12-1.70$ for two two-indexed flavors and three colors with the lower and upper bound set by the minimum and maximum value respectively over all three schemes and at three and four loops. Our analysis suggests that the former two theories lie in the conformal window while the latter belongs to the chirally broken phase. 
 \vskip .1cm
{\footnotesize  \it Preprint:  CP$^3$-Origins-2014-030  DNRF 90\ \& DIAS-2014-30}
 \end{abstract}

\maketitle

\newpage
     
\section{Introduction}

One of the key discoveries that lead to the establishment of the Standard Model of particle physics was the realization that nonabelian gauge theories under a set of appropriate conditions exhibit the phenomena of asymptotic freedom \cite{Gross:1973id,Politzer:1973fx,Gross:1973ju,Gross:1974cs,Politzer:1974fr}. Physically this means that for instance in the case of the strong interactions, described by Quantum Chromo Dynamics (QCD), the quarks behave as if they are free and noninteracting at short distances. 

The discovery is a consequence of a one loop computation of the beta function dictating how the coupling constant changes as one varies the energy scale. The beta function can generically be expanded as a polynomial in the coupling constant and a one loop computation reveals the first coefficient in this expansion to be negative for a sufficiently small number of matter fields. Evaluating the running of the gauge coupling as a function of the energy scale one therefore finds that in the ultraviolet the gauge coupling tends to zero canceling the mutual interaction between the matter fields prompting them to behave freely. On the other hand as the energy is decreased the value of the gauge coupling increases eventually blowing up in the deep infrared invalidating the perturbative expansion. 

It should be clear that this immediately begs the question of what happens if one includes the second term in the perturbative expansion of the beta function. Is the qualitative picture of an ever increasing coupling constant at lower energies, presumably leading to confinement and chiral symmetry breaking, kept intact or could new phenomena and behavior emerge? 

This question was first addressed by Caswell \cite{Caswell:1974gg} and later by Banks and Zaks \cite{Banks:1981nn} who computed and studied the inclusion of the second term in the beta function. Remarkably they found that as they varied the number of matter fields just below where the theory becomes asymptotically free there is a region in which the second coefficient is positive as opposed to the first coefficient which is negative. This has the dramatic effect of turning the beta function around such that it crosses zero for a finite positive value of the coupling constant instead of blowing down to minus infinity. In other words such systems must possess a nontrivial infrared fixed point. As the energy is lowered the coupling constant increases eventually settling at the fixed point value with the theory exhibiting scale invariance. 

As the number of matter fields is lowered even further it was also noted that there was a critical point where the second coefficient of the beta function would eventually become negative similarly to the first coefficient. In this regime there would be no sign of scale invariance. On the contrary the phase would presumably be characterized similarly to what is expected from QCD with confinement, chiral symmetry breaking, massive composite states, etc. 

Now a second question seems to be pressing us. What is the critical number of matter fields where the system transitions from a phase with scale invariance to a phase without scale invariance in the deep infrared? As we tune the number of flavors and approach, from above, the critical value where the second coefficient changes sign the fixed point value of the coupling constant blows to infinity. Again we seem to be haunted by the limits of truncating a perturbative expansion of the beta function.

In the last 30 years there has been a tremendous effort put into the task of estimating this critical value of the number of matter fields where the phase transition occurs. The phase has eventually become known as the conformal window. First, many attempts have been made to estimate the critical value of the coupling constant that is needed to trigger chiral symmetry breaking using truncated Dyson-Schwinger equations \cite{Holdom:1984sk,Yamawaki:1985zg,Appelquist:1986tr,Appelquist:1986an,Appelquist:1988yc,Appelquist:1987fc,Appelquist:1997dc,Brodsky:2008be}. Then by equating the critical value needed for chiral symmetry breaking with the fixed point value stemming from the two loop beta function one can obtain an estimate of the conformal window \cite{Sannino:2004qp,Dietrich:2006cm}. This approach is known as the Ladder approximation.    

A different strategy has been to simulate various theories on a computer and study whether they exhibited scale invariance or not in the deep infrared. In recent years these lattice simulations have received much interests. Investigations have been concerned with a plethora of different $SU(N)$ gauge theories including systems with fundamental matter for two colors \cite{Bursa:2010xr,Bursa:2010xn,Karavirta:2011zg,Appelquist:2013pqa,Hietanen:2013fya,Hietanen:2014xca} and three colors \cite{Appelquist:2007hu,Deuzeman:2008sc,Deuzeman:2009mh,Appelquist:2009ty,Jin:2009mc,Fodor:2009wk,Fodor:2011tu,Appelquist:2011dp,Deuzeman:2012ee,Jin:2012dw}, systems with adjoint matter for two colors \cite{Catterall:2007yx,Hietanen:2008mr,Catterall:2008qk,DelDebbio:2008zf,DelDebbio:2009fd,Bursa:2009we,Hietanen:2009az,DelDebbio:2010hx,DelDebbio:2010hu,Catterall:2011zf,Bursa:2011ru,DeGrand:2011qd,Karavirta:2011mv}, systems with two-indexed symmetric matter for three colors \cite{DeGrand:2008kx,Shamir:2008pb,Fodor:2009ar,Shamir:2010cq,DeGrand:2010na,Fodor:2012ty,DeGrand:2012yq,DeGrand:2013uha} and four colors \cite{DeGrand:2012qa} and finally $SO(4)$ gauge theories with fundamental matter \cite{Hietanen:2012sz,Hietanen:2012qd}. 

In this paper we attempt to estimate the boundary of the conformal window using higher order calculations of the beta function and anomalous dimension of the mass. We perform the analysis in two different explicit schemes. This is the modified  regularization invariant, RI', scheme \cite{Martinelli:1994ty,Gracey:2003yr} and the minimal momentum subtraction scheme mMOM \cite{vonSmekal:2009ae,Celmaster:1979km,Chetyrkin:2000fd,Gracey:2011pf,Gracey:2011vw,Gracey:2013sca}. We then compare our results to earlier similar investigations \cite{Ryttov:2010iz,Pica:2010xq} performed in the usual modified minimal subtraction, $\overline{\text{MS}}$, scheme. The analysis is done at the four loop level and therefore extends earlier investigations done at the three loop level \cite{Ryttov:2013hka,Ryttov:2013ura}. It is the most up-to-date and complete investigation of higher order calculations and their effect on the phase diagram for vectorial fermionic gauge theories. As we shall there is a generic pattern in our computations showing that the anomalous dimension of the mass is smaller at the three and four loop level than the corresponding evaluation at two loops. 

It should be noted that all three schemes are of course related perturbatively in the sense that they give the same predictions at the ultraviolet fixed point. However as has been stressed in \cite{Ryttov:2012ur,Ryttov:2012nt} one can imagine schemes that share this property, i.e. are perturbatively related, but are not suitable for the study of infrared fixed points. For a scheme transformation to be  physically acceptable it must satisfy the following conditions \cite{Ryttov:2012ur,Ryttov:2012nt}: 1) it must map a real positive value of the coupling constant $\alpha$ to a real and positive value of the coupling constant $\alpha'$, 2) it must map a moderate value of $\alpha$ to a moderate value of $\alpha'$, 3) the Jacobian of the scheme transformation should not vanish in the region of $\alpha$ and $\alpha'$ of interest and 4) it must preserve the existence of the infrared zeroes of the beta functions. A priory it is not clear whether the scheme transformations connecting the $\overline{\text{MS}}$, RI' and mMOM schemes possess these properties. In general the conversion functions are highly complicated expressions that depend on the theory under investigation, the gauge parameter, etc, and it is not clear whether they satisfy the above constraints. This provides another reason for our studies.

For completeness and the readers convenience we note that there has also been other orthogonal methods developed in order to calculate the conformal window \cite{Ryttov:2007cx,Pica:2010mt,Antipin:2009wr,Appelquist:1999hr,Appelquist:1999vs}. Also studies of systems with multiple distinct fermion representations \cite{Ryttov:2009yw,Ryttov:2010hs}, Yukawa theories \cite{Molgaard:2014mqa,Molgaard:2014hpa,Antipin:2013sga} and fermions belonging to a spinorial representation or a representation of an exceptional gauge group has been done \cite{Mojaza:2012zd}. Much work trying to clarify the issues concerning scheme dependence can be found in \cite{Shrock:2013uaa,Shrock:2014qea} while additional complimentary investigations into higher order corrections can be found in \cite{Ryttov:2012qu,Shrock:2013ca,Shrock:2013pya,Shrock:2013cca,Shrock:2014zca}.

The paper is organized as follows. In Section \ref{sec:gen} we introduce our calculational setup while Section \ref{sec:results} is devoted to a complete presentation of our numerical results. We finally conclude in Section \ref{sec:conclusion}. Appendix \ref{app:beta} and \ref{app:group} contain all useful relations and quantities used throughout the paper while all of our numerical results and associated plots can be found in Appendix \ref{app:tables} and \ref{app:ComparisonLoopOrders}.

\section{Generalitites}\label{sec:gen}

Our considerations are centered on massless vector-like gauge theories with a set of $N_f$ Dirac fermions transforming according to a representation $r$ of the $G=SU(N)$ gauge group. At the classical level these theories enjoy scale invariance which is spoiled, however, at the quantum level. The quantum fluctuations give rise to a coupling constant which changes (or runs) as one changes the scale. 

Massless vector-like fermionic gauge theories contain two dimensionless parameters. This is the gauge coupling $\alpha$ and the covariant gauge parameter $\xi$. Both of these parameters will receive quantum corrections and scale with the energy. They each have their own associated beta function $\beta_{\alpha}$ and $\beta_{\xi}$ respectively. On the most general grounds one would expect them to be coupled such that each beta function depends on both the gauge coupling and the gauge parameter. Hence we should formally write the renormalization group functions as
\begin{eqnarray}
\mu \frac{d \alpha}{d \mu} &=& \beta_{\alpha} \left( \alpha,\xi \right) \\
\mu \frac{d \xi}{d \mu} &=& \beta_{\xi} \left( \alpha, \xi \right)
\end{eqnarray}
where $\mu$ is the scale. If we want to know the running of the coupling constant we should solve the coupled set of differential equations. In perturbation theory the beta function is an expansion in the coupling constant where in principle we can calculate order by order each term. It is important to remember that the beta function is also a scheme dependent object. By far the most popular scheme in which to calculate the beta function has been the original minimal subtraction, MS, scheme \cite{'tHooft:1973mm} or the modified minimal subtraction, $\overline{\text{MS}}$ scheme \cite{Bardeen:1978yd}. Within these two schemes one enjoys the simplicity that the gauge coupling beta function becomes a function of only the coupling constant. The above two renormalization group equations decouple. Within the class of MS-type schemes \cite{Mojaza:2012mf,Brodsky:2013vpa} it can be shown that the first two coefficients of the beta function are scheme independent. This however is not true on the most general grounds where the gauge coupling beta function also depends on the gauge parameter. Here it is only the first coefficient which is scheme independent. As we will see below this is the case for the mMOM scheme but not the RI' scheme for which the beta function of the gauge coupling actually coincides with the one in the $\overline{\text{MS}}$ scheme. 

If we want to study the fixed points of the theory we need to solve the coupled set of equations 
\begin{eqnarray}
0 &=& \beta_{\alpha} \left( \alpha_0,\xi_0 \right) \\
0 &=& \beta_{\xi} \left( \alpha_0, \xi_0 \right) 
\end{eqnarray}
for the fixed points values $\alpha_0$ and $\xi_0$ of the coupling constant and the gauge parameter. In addition we should classify all the fixed points. At a given loop order there will typically be many solutions to the fixed point equations for which we are only interested in the solutions that are infrared stable with $\alpha_0>0$. At the three loop level all the fixed points were classified in the RI' and mMOM scheme in \cite{Ryttov:2013hka,Ryttov:2013ura}. Here we want to extend the analysis to the four loop order.  

First note that $\beta_{\xi}$ is proportional to $\xi$. This is so since $\beta_{\xi} = \xi \gamma_\xi$ where $\gamma_{\xi}$ is the anomalous dimension of the gauge parameter. Hence we are always guaranteed to find a fixed point solution at $\xi_0=0$ (Landau gauge). In addition if we fix the gauge to the Landau gauge $\xi=0$ in the ultraviolet then as we evolve the renormalization group equation for the gauge parameter it is kept in the Landau gauge $\xi(\mu)=0$ for all energy scales $\mu$. It is then a matter of solving the last fixed point equation $\beta_{\alpha}\left(\alpha_0,0 \right)=0$ in order to find the first nontrivial positive solution $\alpha_0$. If the theory is asymptotically free this solution is then guaranteed to be infrared stable in the $\alpha$ direction. At last we remind the reader that the first coefficient of the beta function of the coupling constant is scheme independent in \emph{any} scheme such that asymptotic freedom is a well defined property.

One should note that typically there will be fixed points for which $\xi_0 \neq 0$ which are also infrared stable. At two and three loops these solutions are either found to reside in smaller ranges of the number of flavors or they fail to predict the correct limit for the anomalous dimension as one approaches the critical number of flavors where asymptotic freedom is lost \cite{Ryttov:2013hka,Ryttov:2013ura}. First in the RI' scheme there are four independent real solutions. One solution is the Landau gauge solution, two solutions only exists for a limited number flavors and one solution does not predict the correct limit for the anomalous dimension. Second in the mMOM scheme there exists two independent real solutions. The first is again the Landau gauge solution while the second generally only exists for a limited number of flavors. In the special case of adjoint fermions the solution exists only at three loops and not at two loops. 

With these considerations in mind we focus our attention at four lops on the Landau gauge solution. Although it would be desirable to do a search for all fixed points at four loops it seems reasonable in the light of the discussion presented above to initially limit ourselves to $\xi_0=0$. At two and three loops in both the RI' and mMOM schemes this gives the only fully consistent and trustable solution.

As has been noted several times the beta function is a scheme dependent quantity and so is the fixed point value $\alpha_0$. Since one of our aims is to compare the predictions of different schemes to a given loop order we need to find a quantity which is scheme independent at the fixed point. This is the anomalous dimension of the mass $\gamma \left( \alpha,\xi \right)$. It has been computed to four loop order in both the RI' scheme \cite{Gracey:2003yr} and the mMOM scheme \cite{Gracey:2013sca} in any gauge covariant gauge. This allows for a comparison with a similar analysis at four loops \cite{Ryttov:2010iz,Pica:2010xq} in the $\overline{\text{MS}}$ scheme \cite{vanRitbergen:1997va,Vermaseren:1997fq}. 

Our setup should now be clear. In the Landau gauge we look for positive solutions to the fixed point equation $\beta_{\alpha} (\alpha_0,0)=0$ and evaluate the associated anomalous dimension of the mass $\gamma(\alpha_0,0)$. From now on we will no longer explicitly write that we are considering Landau gauge $\xi=0$.

\section{Discussion of Results}\label{sec:results}

One can expand both the beta function and the anomalous dimension of the mass in the coupling constant 
\begin{eqnarray}
\beta_{\alpha} \left( \alpha \right) &=& - b_1 \left( \frac{\alpha}{4\pi} \right)^2 - b_2 \left( \frac{\alpha}{4\pi} \right)^3  - b_3 \left( \frac{\alpha}{4\pi} \right)^4 - b_4 \left( \frac{\alpha}{4\pi} \right)^5 + O\left(\alpha^6 \right) \\
\gamma \left( \alpha \right) &=& c_1 \left( \frac{\alpha}{4\pi} \right)  + c_2 \left( \frac{\alpha}{4\pi} \right)^2 +c_3 \left( \frac{\alpha}{4\pi} \right)^3 +c_4 \left( \frac{\alpha}{4\pi} \right)^4 + O\left( \alpha^5 \right)
\end{eqnarray}
The coefficients $b_i$ and $c_i$, $i=1\ldots 4$, have all been computed through four loop order in the $\overline{\text{MS}}$ \cite{vanRitbergen:1997va,Vermaseren:1997fq}, RI' \cite{Gracey:2003yr} and mMOM  \cite{Gracey:2013sca} schemes. They can be found in Appendix \ref{app:beta}. They depend on the number of colors and flavors as well as certain group invariants. We denote by $T(r)$ and $C_2(r)$ the trace normalization factor and quadratic Casimir respectively. These group invariants enter at every loop order At the four loop order yet another set of invariants appear in the coefficients. These are the completely symmetric fourth order tensors $d^{abcd}_r$. For the readers convenience we provide in Appendix \ref{app:group} the explicit expression for all of the group invariants for all the representations used in this work.  This includes the fundamental representation denoted by F, the adjoint representation denoted by G, the two-indexed symmetric representation denoted by 2S and the two-indexed antisymmetric representation denoted by 2A. The dimension of each representation is finally denoted by $d(r)$. At last we note that the coefficients also contain $\zeta_n \equiv \zeta(n)$ which is the Riemann zeta function.

We are searching for infrared fixed points in a range of flavors just below where asymptotic freedom is lost
\begin{eqnarray}
N_f &\leq& \frac{11}{4} \frac{C_2(G)}{T(r)}
\end{eqnarray}
This occurs when the first coefficient of the beta function changes sign. At any loop order the beta function always has a double zero at the origin. This is the ultraviolet fixed point. In addition at two loops there is a single nontrivial zero, at three loops there is a pair of zeros which are the complex conjugate of each other and at four loops there are three nontrivial zeros. In all situations we shall pick the first positive zero away from the origin.

At four loops there is the possibility that the first positive zero away from the origin is not the only infrared stable fixed point. If there exists three distinct real zeros then if the additional two zeros are also positive then the third one away from the origin will be infrared stable while if the additional two zeros are both negative then the second negative zero away from the origin will be infrared stable. The first hypothetical situation with three positive zeros is not observed in any of the three schemes for any of the theories studied here. This fact was first observed in the $\overline{\text{MS}}$ scheme in \cite{Pica:2010xq} but is now seen to occur in both the RI' and mMOM schemes as well. The second situation is observed for certain theories but since the fixed point is at a negative value of the coupling constant these theories must be nonunitary and hence are not included. We conclude that in all three different schemes at four loops and in Landau gauge (where needed) there is only a single physical infrared fixed point. 

There is also a lower bound on the number of flavors that we consider. This is fixed by the value of the coupling constant reaching order one.

In Appendix \ref{app:tables} we provide a comprehensive list of tables all with explicit values of the coupling constant and the anomalous dimension of the mass evaluated at the infrared fixed point in all three different schemes. This is done for the fundamental, adjoint and two-indexed representations and for a range of number of colors and flavors. 

As one decreases the number of flavors below where asymptotic freedom is lost the value of the coupling constant at the infrared fixed point increases. We stop our search for infrared fixed points once the coupling constant reaches a value of order one. We have chosen an exception to this restriction for fermions in the fundamental and two-indexed antisymmetric representations where we have included numerical values of the coupling constant for a certain number of flavors even though $\alpha_0>1$. This is because $\alpha_0$ might be larger than one in certain schemes and at certain loop orders and less than one in other schemes and loop orders. An example of this is the theory with two colors and seven fundamental flavors where the two, three and four loop evaluations in both the $\overline{\text{MS}}$ and RI' schemes and the two loop evaluation in the mMOM scheme predict a value $\alpha_0>1$ while the three and four loop evaluations in the mMOM scheme scheme predict $\alpha_0<1$. In such a case we have included the results in the table and indicated with parenthesis that the evaluation is not trustable since $\alpha_0>1$. In addition if the value of the coupling constant is larger than one then the associated value of the anomalous dimension is equally put in parenthesis to indicate that the computation is no longer trustable.

In addition in Appendix \ref{app:ComparisonLoopOrders} we provide plots of the anomalous dimension as a function of the number of flavors. Each plot shows the two, three and four loop evaluation in a given scheme for different representations and various number of colors. 

In order for a conformal theory not to contain negative normed states the full dimension of any spinless operator must be larger than unity \cite{Mack:1975je,Flato:1983te,Dobrev:1985qv}. In particular it must true for the bilinear $\bar{\psi}\psi$ operator where $\psi$ is the fermion. Therefore since $D(\bar{\psi}\psi)=3-\gamma$ the anomalous dimension must be bounded by $\gamma \leq 2$ in a conformal theory. The plots in Appendix \ref{app:ComparisonLoopOrders} are therefore cut off at this maximal possible value of the anomalous dimension. We stress that the conformal window not necessarily extends all the way to $\gamma=2$. For instance in the Ladder approximation the lower boundary of the conformal window is set by the anomalous dimension reaching the value unity \cite{Holdom:1984sk,Yamawaki:1985zg,Appelquist:1986tr,Appelquist:1986an,Appelquist:1988yc,Appelquist:1987fc,Appelquist:1997dc}. Similarly in supersymmetric QCD the conformal window extends to the point where the anomalous dimension of $\tilde{\Phi}\Phi$ where $\tilde{\Phi}$ and $\Phi$ are chiral superfields reaches unity \cite{Seiberg:1994pq,Ryttov:2007sr}. However by cutting off the plots at $\gamma=2$ we allow for the largest possible value that the anomalous dimension can conceivably take at an infrared fixed point. The plots are also cut off at a number of flavors around the point at which the value of the coupling constant reaches order unity. The solid parts of the curves then correspond to $\alpha_0<1$ while the dashed parts correspond to $\alpha_0>1$.

By examining the different tables and plots the trend should be clear. Both the three and four loop evaluations of the anomalous dimension of the mass is smaller relative to the two loop evaluation. In addition in all but a few special cases the four loop result is lower than the three loop result. In the mMOM scheme for two, three and four colors with fermions in the fundamental representation and for four colors with fermions in the two-indexed antisymmetric representation the four loop evaluation is a bit larger than the three loop evaluation. The trend is seen for all representations, various numbers of colors and in all three different schemes. 

It is important to ask about the reliability of the perturbative calculation. Naively  one might expect that as long as $\alpha_0<1$ then the results can be trusted. This is not always the case however. For fermions in the adjoint and two-indexed symmetric representations the three and four loop evaluation of $\gamma$ in all three schemes are very similar. The same is true for fermions in the fundamental and two-indexed antisymmetric representation in the mMOM scheme. But for fermions in the fundamental representation and the two-indexed antisymmetric representation in the $\overline{\text{MS}}$ and RI' schemes and for a sufficiently small number of flavors, i.e. at the lower end of the conformal window, the behavior of the anomalous dimension differs significantly between the three and four loop evaluation. This occurs even though $\alpha_0<1$ and perturbation theory is expected to be a reasonable approximation. This could be due to the fact that the perturbative expansion has not yet converged.

There are a few theories which are of specific interest to both the lattice and beyond the Standard Model physics communities. First is the theory with twelve fundamental flavors and three colors. Here the three and four loop evaluations of the fixed point coupling constant and anomalous dimension is in quite good agreement among all three schemes with a prediction of $\alpha_0 \sim 0.364 - 0.470$ and $\gamma \sim 0.225-0.375$. Second is the theory with two adjoint flavors and two colors for which the three and four loop evaluations are almost identical. This is again the case in all three different schemes with a prediction $\alpha_0 \sim 0.398 - 0.459$ and $\gamma \sim 0.500 - 0.593$. Third is the theory with two two-indexed symmetric flavors and three colors. At the three and four loop level the prediction of the fixed point value is $\alpha_0 \sim 0.394 - 0.500$ while the anomalous dimension is $\gamma \sim 1.12-1.70$. In this case there is considerable deviation in the value of the anomalous dimension between the various loop orders and schemes. 

This could be a sign of the limitations of the perturbative analysis (poor convergence of the perturbative expansion) and/or be due to the fact that the theory is not within the conformal window. Remember that the anomalous dimension is only a scheme independent quantity at a fixed point. Hence if the theory has not reached the fixed point (and never will) it is no surprise that if one attempts to compute the anomalous dimension in different schemes one will obtain a wide range of different values. The hint that this might have occurred comes from the fact that the anomalous dimension has exceeded unity and according to the Ladder approximation instead must have entered a chirally broken phase. However we warn the reader that in order to fully access whether the theory is inside or outside the conformal window requires complete nonperturbative knowledge about its dynamics.

It should also be noted that the former two theories with twelve fundamental flavors and three colors or two adjoint flavors and two colors lie very close to the point where deviations between the three and four loop evaluations in certain schemes start to become significant. Similarly this could be a sign that these theories lie very close to (but within) the boundary of the conformal window.

\section{Conclusion}\label{sec:conclusion}

We have searched for infrared fixed points in theories with massless vector-like fermionic matter and gauge group $SU(N)$. This was done by analyzing the beta function of the coupling constant and the anomalous dimension of the mass at two, three and four loop level. The search was done in the RI' and mMOM schemes and then compared to earlier studies in the $\overline{\text{MS}}$ scheme. In cases where the beta function depended on the gauge parameter the investigations were performed in the Landau gauge.

We found a generic pattern with the value of the anomalous dimension at three and four loops being smaller as compared to its two loop evaluation. This was found in all three different schemes. We then discussed specifically the prediction of the anomalous dimension for three different theories. This was twelve fundamental flavors and three colors, two adjoint flavors and two colors and two two-indexed symmetric flavors and three colors. In the first two models our prediction of the anomalous dimension was $\gamma \sim 0.225-0.375$ and $\gamma \sim 0.500 - 0.593$ respectively while in the third model it was $\gamma \sim 1.12-1.70$. This seems to indicate that the first two theories belong to the conformal window while the third belongs to the chirally broken phase. 

\appendix

\section{Beta Function and Anomalous Dimension of the Mass to Four Loops}\label{app:beta}

\subsection{$\overline{\text{MS}}$ Scheme}

Coefficients in the modified minimal subtraction, $\overline{\text{MS}}$, scheme of the beta function and the anomalous dimension of the mass to four loop order
\begin{eqnarray}
b_1 &=& \frac{11}{3} C_2(G) - \frac{4}{3} T(r) N_f \\
b_2 &=& \frac{34}{3} C_2(G)^2 - 4 C_2(r)T(r)N_f - \frac{20}{3} C_2(G) T(r) N_f \\
b_3 &=&   \frac{2857}{54} C_2(G)^3  +2 C_2(r)^2  T(r) N_f -\frac{205}{9} C_2(r) C_2(G) T(r) N_f - \frac{1415}{27} C_2(G)^2 T(r) N_f \nonumber \\
&& + \frac{44}{9} C_2(r) T(r)^2 N_f^2 + \frac{158}{27} C_2(G) T(r)^2 N_f^2 \\
b_4 &=&  \left( \frac{150653}{486} -\frac{44}{9}\zeta_3  \right)C_2(G)^4 +  \left( - \frac{39143}{81} + 
\frac{136}{3} \zeta_3 \right)C_2(G)^3 T(r) N_f \nonumber \\
&&   +  \left( \frac{7073}{243} - \frac{656}{9} \zeta_3\right)C_2(G)^2 C_2(r) T(r)N_f +  \left(-\frac{4204}{27} + \frac{352}{9} \zeta_3 \right)C_2(G) C_2(r)^2 T(r) N_f \nonumber \\
&& +46 C_2(r)^3 T(r) N_f +   \left(\frac{7930}{81} + \frac{224}{9} \zeta_3 \right)C_2(G)^2 T(r)^2 N_f^2 + \left(\frac{1352}{27} - \frac{704}{9} \zeta_3 \right) C_2(r)^2 T(r)^2 N_f^2  \nonumber \\
&& +  \left( \frac{17152}{243} + \frac{448}{9}  \zeta_3 \right)C_2(G) C_2(r) T(r)^2 N_f^2 + \frac{424}{243} C_2(G) T(r)^3 N_f^3 + \frac{1232}{243}C_2(r) T(r)^3 N_f^3 \nonumber \\
&& +  \left( -\frac{80}{9} + \frac{704}{3} \zeta_3 \right) \frac{d_G^{abcd} d_G^{abcd}}{d(G)}+ \left(  \frac{512}{9} - \frac{1664}{3}\zeta_3 \right)N_f \frac{d_r^{abcd}d_G^{abcd}}{d(G)} \nonumber \\
&& +  \left(- \frac{704}{9} + \frac{512}{3} \zeta_3 \right)N_f^2 \frac{d_r^{abcd} d_r^{abcd}}{d(G)}
\end{eqnarray}

\begin{eqnarray}
c_1  &=& 6 C_2(r) \\
c_2  &=& 3 C_2(r)^2  + \frac{97}{3} C_2(G) C_2(r) - \frac{20}{3} C_2(r) T(r) N_f \\
c_3 &=& 129 C_2(r)^3 - \frac{129}{2} C_2(G) C_2(r)^2 + \frac{11413}{54} C_2(G)^2 C_2(r) + \left( -92 + 96 \zeta_3 \right) C_2(r)^2 T(r)N_f \nonumber \\
&& -\left( \frac{1112}{27} + 96 \zeta_3 \right) C_2(G) C_2(r) T(r) N_f - \frac{280}{27} C_2(r) T(r)^2 N_f^2 \\
c_4 &=& - \left( \frac{1261}{4} + 672 \zeta_3 \right) C_2(r)^4 + \left(\frac{15349}{6} +632 \zeta_3 \right) C_2(G) C_2(r)^3 \nonumber \\
&& + \left(- \frac{34045}{18} - 304 \zeta_3 + 880 \zeta_5  \right) C_2(G)^2 C_2(r)^2 + \left( \frac{70055}{36} + \frac{2836}{9} \zeta_3 - 880 \zeta_5 \right) C_2(G)^3 C_2(r) \nonumber \\
&& + \left( - \frac{560}{3} + 1104 \zeta_3 - 960 \zeta_5 \right) C_2(r)^3 T(r)N_f + \left( - \frac{17638}{27} + 736 \zeta_3 - 528 \zeta_4 + 160 \zeta_5 \right) C_2(G) C_2(r)^2 T(r)N_f \nonumber \\
&& + \left( - \frac{65459}{81} - \frac{5368}{3}\zeta_3 + 528 \zeta_4 + 800 \zeta_5 \right) C_2(G)^2 C_2(r) T(r) N_f \nonumber \\
&& + \left( \frac{608}{27} - 320 \zeta_3 + 192\zeta_4 \right) C_2(r)^2T(r)^2 N_f^2 + \left( \frac{2684}{81} + 320 \zeta_3 - 192 \zeta_4 \right) C_2(G) C_2(r) T(r)^2 N_f^2 \nonumber \\
&& + \left( -\frac{1328}{81} + \frac{256}{9}\zeta_3 \right) C_2(r) T(r)^3 N_f^3 + \left(-64 + 480 \zeta_3  \right) \frac{d_r^{abcd}d_G^{abcd}}{d(r)} + \left( 128 - 960\zeta_3 \right) N_f \frac{d_r^{abcd}d_r^{abcd}}{d(r)}
\end{eqnarray}

\subsection{RI' Scheme }

Coefficients in the modified regularization invariant, RI', scheme of the beta function and the anomalous dimension of the mass to four loop order and in Landau gauge $\xi=0$
\begin{eqnarray}
b_1 &=& \frac{11}{3} C_2(G) - \frac{4}{3} T(r) N_f \\
b_2 &=& \frac{34}{3} C_2(G)^2 - 4 C_2(r)T(r)N_f - \frac{20}{3} C_2(G) T(r) N_f \\
b_3 &=&   \frac{2857}{54} C_2(G)^3  +2 C_2(r)^2  T(r) N_f -\frac{205}{9} C_2(r) C_2(G) T(r) N_f - \frac{1415}{27} C_2(G)^2 T(r) N_f \nonumber \\
&& + \frac{44}{9} C_2(r) T(r)^2 N_f^2 + \frac{158}{27} C_2(G) T(r)^2 N_f^2 \\
b_4 &=& C_2(G)^4 \left( \frac{150653}{486} -\frac{44}{9}\zeta_3  \right) + C_2(G)^3 T(r) N_f \left( - \frac{39143}{81} + 
\frac{136}{3} \zeta_3 \right) \nonumber \\
&&   + C_2(G)^2 C_2(r) T(r)N_f \left( \frac{7073}{243} - \frac{656}{9} \zeta_3\right) + C_2(G) C_2(r)^2 T(r) N_f \left(-\frac{4204}{27} + \frac{352}{9} \zeta_3 \right) \nonumber \\
&& +46 C_2(r)^3 T(r) N_f + C_2(G)^2 T(r)^2 N_f^2  \left(\frac{7930}{81} + \frac{224}{9} \zeta_3 \right) + C_2(r)^2 T(r)^2 N_f^2 \left(\frac{1352}{27} - \frac{704}{9} \zeta_3 \right) \nonumber \\
&& + C_2(G) C_2(r) T(r)^2 N_f^2 \left( \frac{17152}{243} + \frac{448}{9}  \zeta_3 \right) + \frac{424}{243} C_2(G) T(r)^3 N_f^3 + \frac{1232}{243}C_2(r) T(r)^3 N_f^3 \nonumber \\
&& + \frac{d_G^{abcd} d_G^{abcd}}{d(G)} \left( -\frac{80}{9} + \frac{704}{3} \zeta_3 \right) + N_f \frac{d_r^{abcd}d_G^{abcd}}{d(G)}\left(  \frac{512}{9} - \frac{1664}{3}\zeta_3 \right) \nonumber \\
&& + N_f^2 \frac{d_r^{abcd} d_r^{abcd}}{d(G)} \left(- \frac{704}{9} + \frac{512}{3} \zeta_3 \right)
\end{eqnarray}

\begin{eqnarray}
c_1 &=& 6 C_2(r) \\
c_2 &=& \frac{1}{3} \left[  185 C_2(G) + 9 C_2(r) - 52T(r) N_f \right]C_2(r) \\
c_3 &=& \frac{1}{108} \left[ \left( 1944 + 19008\zeta_3  \right) C_2(G) C_2(r) + \left( 117428 - 28512 \zeta_3 \right) C_2(G)^2 \right.  \nonumber \\
&& \left. - 62960 C_2(G) T(r)N_f + 13932 C_2(r)^2 - \left( 16632- 3456\zeta_3 \right) C_2(r) T(r) N_f  + 6848 T(r)^2 N_f^2 \right] C_2(r) \\
c_4 &=& - \frac{1}{3888} \left[  \left( -92569118 + 39004740 \zeta_3 - 1710720 \zeta_5 \right) C_2(G)^3 C_2(r) \right. \nonumber \\
&& + \left( 10355148 - 22203072\zeta_3 - 1710720 \zeta_5  \right) C_2(G)^2 C_2(r)^2 \nonumber \\
&&+  \left( 73217928 - 14239152\zeta_3 - 1244160\zeta_5 \right) C_2(G)^2 C_2(r) T(r)N_f \nonumber \\
&& + \left( -33960384 - 1601856 \zeta_3 + 10264320 \zeta_5  \right) C_2(G) C_2(r)^3 \nonumber \\
&& + \left(20983248 + 1347840 \zeta_3 - 1244160 \zeta_5 \right) C_2(G) C_2(r)^2 T(r)N_f \nonumber \\
&& + \left( -16599552 + 580608 \zeta_3 \right) C_2(G) C_2(r) T(r)^2 N_f^2 + \left( 9745920 - 3856896 \zeta_3 \right) C_2(r)^3 T(r)N_f \nonumber \\
&& + \left( -6653952 + 2571264\zeta_3 \right) C_2(r)^2 T(r)^2 N_f^2 + \left(1225692 + 2612736\zeta_3 \right) C_2(r)^4 \nonumber \\
&& \left. + 1025536 C_2(r) T(r)^3 N_f^3 + \left( 248832 - 1866240 \zeta_3 \right) \frac{d_r^{abcd} d_G^{abcd}}{d(r)} + \left( -497664 + 3732480\zeta_3 \right)N_f \frac{d_r^{abcd}d_r^{abcd}}{d(r)}  \right] \nonumber \\
\end{eqnarray}

\subsection{mMOM Scheme}

Coefficients in the minimal momentum subtraction, mMOM, scheme of the beta function and the anomalous dimension of the mass to four loop order and in Landau gauge $\xi=0$
\begin{eqnarray}
b_1 &=& \frac{11}{3} C_2(G) -\frac{4}{3} T(r) N_f \\
b_2 &=& \frac{34}{3} C_2(G)^2 - 4 C_2(r) T(r)N_f - \frac{20}{3}C_2(G) T(r)N_f \\
b_3 &=& - \frac{1}{288}\left[ \left(- 38620 + 5148 \zeta_3 \right) C_2(G)^3  + \left( 32144 + 6576\zeta_3 \right) C_2(G)^2 T(r)N_f \right. \nonumber \\
&& + \left( 20512 - 16896\zeta_3 \right) C_2(G) C_2(r) T(r) N_f - \left(4416 + 3072\zeta_3 \right) C_2(G) T(r)^2 N_f^2 - 576 C_2(r)^2 \nonumber \\
&&\left.  + \left( - 5888 + 6144\zeta_3 \right) C_2(r)T(r)^2 N_f^2 \right] \\
b_4 &=& - \frac{1}{10368} \left[ \left( -22106704 + 5509416 \zeta_3 + 3090780 \zeta_5 \right) C_2(G)^4 \right. \nonumber \\
&& + \left( 23501280 - 1217376 \zeta_3 - 5178960 \zeta_5 \right) C_2(G)^3 T(r)N_f \nonumber \\
&& + \left(17477280 - 7050240 \zeta_3 - 6082560 \zeta_5 \right) C_2(G)^2 C_2(r) T(r)N_f \nonumber \\
&& + \left( -5719680 - 1654272\zeta_3 + 1474560 \zeta_5 \right) C_2(G)^2 T(r)^2 N_f^2 \nonumber \\
&& + \left( - 607104 - 7907328 \zeta_3 + 12165120 \zeta_5 \right) C_2(G) C_2(r)^2 T(r)N_f \nonumber \\
&& + \left(-10861056 + 4755456\zeta_3 + 2211840 \zeta_5 \right) C_2(G)C_2(r) T(r)^2 N_f^2 \nonumber \\
&& + \left( 229376 + 344064\zeta_3 \right) C_2(G) T(r)^3N_f^3 - 476928 C_2(r)^3 T(r)N_f \nonumber \\
&& + \left( 267264 + 3538944 \zeta_3 - 4423680 \zeta_5 \right) C_2(r)^2 T(r)^2 N_f^2 + \left( 1327104 - 884736 \zeta_3 \right) C_2(r) T(r)^3 N_f^3 \nonumber \\
&& + \left( 92160 - 2433024\zeta_3 \right) \frac{d_G^{abcd}d_G^{abcd}}{d(G)} + \left(-589824 + 5750784 \zeta_3 \right) N_f \frac{d_r^{abcd}d_G^{abcd}}{d(G)} \nonumber \\
&& \left. + \left( 811008 - 1769 472 \zeta_3 \right) N_f^2 \frac{d_r^{abcd}d_r^{abcd}}{d(G)} \right]
\end{eqnarray}

\begin{eqnarray}
c_1 &=& 6 C_2(r) \\
c_2 &=&\frac{1}{2} \left[ 6 C_2(r) + 67 C_2(G) - 8 T(r)N_f \right] C_2(r) \\
c_3 &=& - \frac{1}{24} \left[ \left( -10095 + 5634 \zeta_3 \right) C_2(G)^2 + \left( 244 - 4224 \zeta_3 \right) C_2(G) C_2(r) \right. \nonumber \\
&& + \left( 3888 - 1152 \zeta_3 \right) C_2(G) T(r) N_f - 3096 C_2(r)^2 + \left( 736 + 1536 \zeta_3\right) C_2(r) T(r)N_f \nonumber \\
&& \left. - 384 T(r)^2 N_f^2 \right] C_2(r) \\
c_4 &=& - \frac{1}{2592} \left[\left( -10139319 + 16036470 \zeta_3 - 6334605 \zeta_5  \right) C_2(G)^3 C_2(r) \right. \nonumber \\
&& + \left( -2188530 - 10029096 \zeta_3 + 3421440 \zeta_5 \right) C_2(G)^2 C_2(r)^2 \nonumber \\
&& + \left(8403640 - 15748128\zeta_3 + 2737152 \zeta_4 + 4147200 \zeta_5  \right) C_2(G)^2 C_2(r) T(r)N_f \nonumber \\
&& + \left( -4669704 + 2208384\zeta_3 + 6842880 \zeta_5 \right) C_2(G) C_2(r)^3 \nonumber \\
&& + \left( -2214048 + 7091712\zeta_3 - 2737152 \zeta_4 + 1244160 \zeta_5 \right) C_2(G) C_2(r)^2 T(r) N_f \nonumber \\
&& + \left( -2128192 + 2405376\zeta_3 - 995328 \zeta_4 \right) C_2(G) C_2(r) T(r)^2N_f^2 \nonumber \\
&& + \left(-817128 - 1741824 \zeta_3 \right) C_2(r)^4 + \left( 3509568 + 4935168 \zeta_3 - 7464960 \zeta_5 \right) C_2(r)^3 T(r) N_f \nonumber \\
&& + \left( -605568 - 1327104 \zeta_3 + 995328\zeta_4 \right) C_2(r)^2 T(r)^2 N_f^2 + \left( -2048 + 147456 \zeta_3\right) C_2(r) T(r)^3 N_f^3 \nonumber \\
&& +\left( -165888 + 1244160 \zeta_3 \right) \frac{d_r^{abcd}d_G^{abcd}}{d(r)} + \left( 331776 - 2488320 \zeta_3 \right) N_f\frac{d_r^{abcd}d_r^{abcd}}{d(r)}
\end{eqnarray}

\section{Group Invariants for $SU(N)$}\label{app:group}

\begin{center}
  \begin{tabular}{ c | c | c | c | c | c}
    $r$ & $d(r)$ & $T(r)$ & $C_2(r)$ & $d^{abcd}_r d^{abcd}_{G} $ & $d^{abcd}_r d^{abcd}_{r} $  \\ \hline
    F & $N$ & $\frac{1}{2}$ & $\frac{N^2-1}{2N}$ & $\frac{N(N^2-1)(N^2 +6)}{48}$ & $\frac{(N^2-1)\left( N^4 -6N^2 +18 \right)}{96N^2}$   \\
    G & $N^2 -1$ & $N$ & $N$ & $\frac{N^2 \left( N^2 -1 \right) \left( N^2 + 36 \right)}{24}$ & $\frac{N^2 \left( N^2 -1 \right) \left( N^2 + 36 \right)}{24}$ \\
    2S & $\frac{N(N+1)}{2}$ & $\frac{N+2}{2}$ & $\frac{(N-1)(N+2)}{N}$ & $\frac{N(N^2-1)(N+2)\left(N^2 + 6N +24 \right) }{48}$ & $\frac{(N^2-1)(N+2)\left( N^5 +14 N^4 + 72 N^3 - 48 N^2 - 288 N + 576 \right)}{96N^2}$ \\
   2A &  $\frac{N(N-1)}{2}$ & $\frac{N-2}{2}$ & $\frac{(N+1)(N-2)}{N}$ & $\frac{N(N^2-1)(N-2)\left(N^2 - 6N +24 \right) }{48}$ & $\frac{(N^2-1)(N-2)\left( N^5 -14 N^4 + 72 N^3 + 48 N^2 - 288 N - 576 \right)}{96N^2}$ \\ 
    \hline
  \end{tabular}
\end{center}
 
\clearpage

\section{Values of the Coupling Constant and Anomalous Dimension of the Mass at Fixed Points}\label{app:tables}

\begin{table}[h]
\begin{center}\label{tab:coupling-Fundamental}
  \begin{tabular}{|cc||c|c|c||c|c|c||c|c|c|}
  \hline\hline
  \multicolumn{2}{|c||}{{\color{rossoCP3}  \rm F}}  &   \multicolumn{3}{c||}{$\overline{\text{MS}}$} & \multicolumn{3}{c||}{RI'} & \multicolumn{3}{c|}{mMOM}  \\
\hline
  \hline
    $N$ & $N_f$ & $\alpha_{2l}$ & $\alpha_{3l}$  & $\alpha_{4l}$  & $\alpha_{2l}$  & $\alpha_{3l}$  & $\alpha_{4l}$ & $\alpha_{2l}$ & $\alpha_{3l}$  & $\alpha_{4l}$    \\ \hline
    2 &  6 & (11.4) & (1.65) & (2.40) &(11.4)&(1.65)& (2.40) &(11.4) &(1.26) &(1.11)\\
    2 &  7 & (2.83) & (1.05) & (1.21)  & (2.83) & (1.05) & (1.21)  & (2.83) & 0.854 & 0.790 \\ 
    2 &  8 & (1.26) & 0.688 & 0.760 &(1.26) & 0.688 & 0.760 &(1.26) & 0.588 & 0.571 \\
    2 &  9 & 0.595 &0.418  & 0.444   &0.595 &0.418  & 0.444& 0.595 &0.377 &  0.377 \\
    2 &  10 & 0.231 & 0.196 & 0.200 &0.231 & 0.196 & 0.200 &0.231 &0.187 & 0.188  \\
    \hline
    3 & 9 & (5.24)&(1.03) &(1.07) &(5.24) & (1.03)& (1.07) &(5.24)& 0.810& 0.690\\
    3 &  10 & (2.21) & 0.764 & 0.815 & (2.21) & 0.764 & 0.815 &(2.21)&0.621  &0.556 \\ 
    3 &  11 & (1.23) & 0.579 & 0.626 &(1.23) & 0.579 & 0.626&(1.23)& 0.485 & 0.453 \\
    3 &  12 & 0.754 & 0.435 &0.470 & 0.754 & 0.435 &0.470 &0.754&0.377   &0.364 \\
    3 &  13 & 0.468 &0.317  & 0.337 & 0.468 &0.317  & 0.337 &0.468&0.283   & 0.281 \\
    3 &  14 & 0.278 &0.215  & 0.224 & 0.278 &0.215  & 0.224 &0.278& 0.198  &0.199 \\
    3 &  15 & 0.143 & 0.123 & 0.126 & 0.143 & 0.123 & 0.126 &0.143&0.118   & 0.119 \\
    3 &  16 & 0.0416 & 0.0397& 0.0398 &0.0416 & 0.0397& 0.0398 &0.0416& 0.0392  & 0.0392 \\
    \hline
    4 &  12 &(3.54)&0.754&0.759&(3.54)&0.754&0.759&(3.54)&0.600&0.507\\
    4 &  13 & (1.85) & 0.604 &0.628  & (1.85) & 0.604 &0.628 & (1.85) & 0.490 & 0.432 \\ 
    4 &  14 &(1.16) &0.489& 0.521  &(1.16) &0.489& 0.521 &(1.16) & 0.406& 0.371 \\
    4 &  15 & 0.783 & 0.397 &0.428 & 0.783 & 0.397 &0.428 &0.783 & 0.338 & 0.318 \\
    4 &  16 &0.546 &0.320 & 0.345 & 0.546 &0.320 & 0.345 &0.546 & 0.278 & 0.269 \\
    4 &  17 &0.384 & 0.254 & 0.271 & 0.384 & 0.254 & 0.271 & 0.384 & 0.226 &0.223 \\
    4 &  18 & 0.266& 0.194 & 0.205 & 0.266& 0.194 & 0.205 &0.266& 0.177 &0.177 \\
    4 &  19 & 0.175 & 0.140 & 0.145 & 0.175 & 0.140 & 0.145 &0.175 & 0.131 & 0.132 \\
    4 &  20 &0.105 &0.0907 & 0.0924 & 0.105 &0.0907 & 0.0924 & 0.105&0.0868 &0.0873 \\
    4 &  21 & 0.0472 & 0.0441 & 0.0444 &0.0472 & 0.0441 & 0.0444 & 0.0472 &0.0432 & 0.0433 \\
    \hline
  \end{tabular}
  \caption{Values of the coupling constant at the infrared fixed point at two, three and four loop level in the $\overline{\text{MS}}$, RI' and mMOM schemes. The fermions are in the fundamental representation of the $SU(N)$ gauge group.}
\end{center}
\end{table}

\begin{table}[h]
\begin{center}\label{tab:gamma-Fundamental}
  \begin{tabular}{|cc||c|c|c||c|c|c||c|c|c|}
  \hline\hline
\multicolumn{2}{|c||}{{\color{rossoCP3}  \rm F}}  &   \multicolumn{3}{c||}{$\overline{\text{MS}}$} & \multicolumn{3}{c||}{RI'} & \multicolumn{3}{c|}{mMOM}  \\
\hline
  \hline
    $N$ & $N_f$ & $\gamma_{2l}$ & $\gamma_{3l}$  & $\gamma_{4l}$  & $\gamma_{2l}$  & $\gamma_{3l}$  & $\gamma_{4l}$ & $\gamma_{2l}$ & $\gamma_{3l}$  & $\gamma_{4l}$    \\ 
    \hline
    2 &  6 &(33.2)&(0.925)&(-4.02)&(49.7)&(2.06)&(-0.297)&(39.6)&(1.03)&(1.39)\\
    2 &  7 &(2.67) & (0.457) & (0.0325) &(3.49) &(0.671) & (-0.0227) & (3.12) & 0.523 & 0.628 \\ 
    2 &  8 &(0.752) & 0.272 & 0.204 & (0.872)&0.312 &0.163  & (0.849) & 0.300 &0.338  \\
    2 &  9 & 0.275 & 0.161 & 0.157 & 0.293 &0.166  & 0.152 & 0.299 & 0.169 & 0.179 \\
    2 &  10 & 0.0910&0.0738  & 0.0748 & 0.0924 & 0.0740 &0.0746 &0.0950  &0.0748 &0.0759  \\
    \hline
    3 &   9 &(19.8)&(1.06)&(-0.143)&(29.0)&(2.23)&(1.46)&(23.4)&1.19&1.41 \\
    3 &  10 & (4.19) & 0.647&0.156 & (5.62) &1.04 &0.342  & (4.88) & 0.735 & 0.850 \\ 
    3 &  11 &(1.61) & 0.439&0.250 & (1.99) & 0.572 & 0.221 & (1.85) &0.492 &0.558  \\
    3 &  12 &0.773 &0.312 & 0.253 & 0.888& 0.354& 0.225 & 0.866 &0.340  & 0.375  \\
    3 &  13 &0.404 & 0.220 & 0.210 &0.439 &0.232 & 0.120 & 0.443 & 0.233 &0.249 \\
    3 &  14 & 0.212 & 0.146 &0.147  & 0.221 &0.149  & 0.145 & 0.227 & 0.151 & 0.157 \\
    3 &  15 & 0.0997 & 0.0826 & 0.0836 & 0.101 & 0.0828 &0.0834 & 0.104 & 0.0835 &0.0847 \\
    3 &  16 & 0.0272 & 0.0258 & 0.0259 & 0.0272 &0.0258 & 0.0259 &0.0276  &0.0259 &0.0259 \\
    \hline
    4 &  12 &(17.3) &1.11&0.0584&(25.2)&2.28&1.56&(20.4)&1.24&1.43 \\
    4 &  13 & (5.38) &0.755 &0.192  &(7.33) & 1.27 & 0.558 &(6.28) & 0.856 &0.978  \\ 
    4 &  14 & (2.45) & 0.552 & 0.259 & (3.13) & 0.784 &0.301 &(2.82) & 0.622 &0.706 \\
    4 &  15 & 1.32 &0.420 &0.281 & 1.59 & 0.523 &0.253 & 1.50 &0.466  & 0.522 \\
    4 &  16 & 0.778 &0.325 & 0.269 &0.892 &0.368  &0.243 & 0.871 & 0.354 & 0.388 \\
    4 &  17 & 0.481 &0.251 &0.234 & 0.528 & 0.267 & 0.221 & 0.529 &0.267 & 0.287 \\
    4 &  18 & 0.301 &0.189 & 0.187 &0.318 & 0.194 & 0.182 &0.325 &0.197 & 0.207 \\
    4 &  19 &0.183 & 0.134 & 0.136 &0.189  &0.136 &0.135 & 0.194 & 0.138 & 0.142\\
    4 &  20 & 0.102 & 0.0854& 0.0865 & 0.104 & 0.0856& 0.0863& 0.106 & 0.0864 &0.0875 \\
    4 &  21 & 0.0440 &0.0407 & 0.0409 & 0.0441 & 0.0407 &0.0409 & 0.0449 & 0.0408 & 0.0409 \\
    \hline
  \end{tabular}
  \caption{Values of the anomalous dimension of the mass at the infrared fixed point at two, three and four loop level in the $\overline{\text{MS}}$, RI' and mMOM schemes. The fermions are in the fundamental representation of the $SU(N)$ gauge group.}
\end{center}
\end{table}

\begin{table}[h]
\begin{center}\label{tab:coupling-Adjoint}
  \begin{tabular}{|cc||c|c|c||c|c|c||c|c|c|}
  \hline\hline
\multicolumn{2}{|c||}{{\color{rossoCP3}  \rm G}}  &   \multicolumn{3}{c||}{$\overline{\text{MS}}$} & \multicolumn{3}{c||}{RI'} & \multicolumn{3}{c|}{mMOM}  \\
\hline
  \hline
    $N$ & $N_f$ & $\alpha_{2l}$ & $\alpha_{3l}$  & $\alpha_{4l}$  & $\alpha_{2l}$  & $\alpha_{3l}$  & $\alpha_{4l}$ & $\alpha_{2l}$ & $\alpha_{3l}$  & $\alpha_{4l}$    \\ \hline
    2 &  2 & 0.628 & 0.459 & 0.450 & 0.628 & 0.459 & 0.450 & 0.628 &0.424 &0.398  \\ 
    3 &  2 & 0.419 & 0.306 & 0.308 & 0.419 & 0.306 & 0.308 &0.419 & 0.283 &  0.270 \\
    4 &  2 & 0.314 &0.229  & 0.234 & 0.314 &0.229  & 0.234 & 0.314&0.212 &  0.204 \\
   \hline
  \end{tabular}
  \caption{Values of the coupling constant at the infrared fixed point at two, three and four loop level in the $\overline{\text{MS}}$, RI' and mMOM schemes. The fermions are in the adjoint representation of the $SU(N)$ gauge group.}
\end{center}
\end{table}

\begin{table}[h]
\begin{center}\label{tab:gamma-Fundamental}
  \begin{tabular}{|cc||c|c|c||c|c|c||c|c|c|}
  \hline\hline
\multicolumn{2}{|c||}{{\color{rossoCP3}  \rm G}}  &   \multicolumn{3}{c||}{$\overline{\text{MS}}$} & \multicolumn{3}{c||}{RI'} & \multicolumn{3}{c|}{mMOM}  \\
\hline
  \hline
    $N$ & $N_f$ & $\gamma_{2l}$ & $\gamma_{3l}$  & $\gamma_{4l}$  & $\gamma_{2l}$  & $\gamma_{3l}$  & $\gamma_{4l}$ & $\gamma_{2l}$ & $\gamma_{3l}$  & $\gamma_{4l}$    \\ \hline
    2 &  2 & 0.820 & 0.543 & 0.500 & 0.900 & 0.593 &0.518  & 0.885 & 0.569 & 0.559  \\ 
    3 &  2 & 0.820 & 0.543 & 0.523 & 0.900 & 0.593 &0.541 & 0.885 & 0.569 & 0.568  \\
    4 &  2 & 0.820 & 0.543 & 0.532 & 0.900 & 0.593 & 0.550 &0.885 & 0.569 & 0.571 \\
    \hline
      \end{tabular}
  \caption{Values of the anomalous dimension of the mass at the infrared fixed point at two, three and four loop level in the $\overline{\text{MS}}$, RI' and mMOM schemes. The fermions are in the adjoint representation of the $SU(N)$ gauge group.}
\end{center}
\end{table}

\begin{table}[h]
\begin{center}\label{tab:coupling-Symmetric}
  \begin{tabular}{|cc||c|c|c||c|c|c||c|c|c|}
  \hline\hline
\multicolumn{2}{|c||}{{\color{rossoCP3}  \rm 2S}}  &   \multicolumn{3}{c||}{$\overline{\text{MS}}$} & \multicolumn{3}{c||}{RI'} & \multicolumn{3}{c|}{mMOM}  \\
\hline
  \hline
    $N$ & $N_f$ & $\alpha_{2l}$ & $\alpha_{3l}$  & $\alpha_{4l}$  & $\alpha_{2l}$  & $\alpha_{3l}$  & $\alpha_{4l}$ & $\alpha_{2l}$ & $\alpha_{3l}$  & $\alpha_{4l}$    \\ \hline
    3 &  2  & 0.842 &0.500 &0.470 & 0.842 &0.500 &0.470 & 0.842  &0.460 & 0.394 \\ 
    3 &  3  & 0.0849 &0.0790  & 0.0795 & 0.0849 &0.0790  & 0.0795 &0.0849  & 0.0771 &0.0771 \\
    \hline
    4 &  2  & 0.967 & 0.485 & 0.440& 0.967 & 0.485 & 0.440 &0.967 & 0.451 & 0.358 \\
    4 &  3  & 0.152 & 0.129 &0.131 & 0.152 & 0.129 &0.131 &0.152 & 0.123 & 0.122 \\
   \hline
  \end{tabular}
  \caption{Values of the coupling constant at the infrared fixed point at two, three and four loop level in the $\overline{\text{MS}}$, RI' and mMOM schemes. The fermions are in the two-indexed symmetric representation of the $SU(N)$ gauge group.}
\end{center}
\end{table}

\begin{table}[h]
\begin{center}\label{tab:gamma-Symmetric}
  \begin{tabular}{|cc||c|c|c||c|c|c||c|c|c|}
  \hline\hline
\multicolumn{2}{|c||}{{\color{rossoCP3}  \rm 2S}}  &   \multicolumn{3}{c||}{$\overline{\text{MS}}$} & \multicolumn{3}{c||}{RI'} & \multicolumn{3}{c|}{mMOM}  \\
\hline
  \hline
    $N$ & $N_f$ & $\gamma_{2l}$ & $\gamma_{3l}$  & $\gamma_{4l}$  & $\gamma_{2l}$  & $\gamma_{3l}$  & $\gamma_{4l}$ & $\gamma_{2l}$ & $\gamma_{3l}$  & $\gamma_{4l}$    \\ \hline
    3 &  2  & 2.44 &1.28  & 1.12 & 2.96 &  1.70 &1.55 &2.69 &1.42 &1.26  \\ 
    3 &  3  & 0.144 &0.133 & 0.133 &0.145 &  0.133 & 0.133& 0.147& 0.133&0.134 \\
    \hline
    4 &  2  & 4.82 & 2.08 &1.79  & 6.24 & 3.19 & 3.30 &5.37 & 2.44 & 1.93 \\
    4 &  3  & 0.381 &0.313 &0.315  & 0.395 & 0.319 &0.316 & 0.400 &0.319 &0.321  \\
    \hline
      \end{tabular}
  \caption{Values of the anomalous dimension of the mass at the infrared fixed point at two, three and four loop level in the $\overline{\text{MS}}$, RI' and mMOM schemes. The fermions are in the two-indexed symmetric representation of the $SU(N)$ gauge group.}
\end{center}
\end{table}

\begin{table}[h]
\begin{center}\label{tab:coupling-Antisymmetric}
  \begin{tabular}{|cc||c|c|c||c|c|c||c|c|c|}
  \hline\hline
\multicolumn{2}{|c||}{{\color{rossoCP3}  \rm 2A}}  &   \multicolumn{3}{c||}{$\overline{\text{MS}}$} & \multicolumn{3}{c||}{RI'} & \multicolumn{3}{c|}{mMOM}  \\
\hline
  \hline
    $N$ & $N_f$ & $\alpha_{2l}$ & $\alpha_{3l}$  & $\alpha_{4l}$  & $\alpha_{2l}$  & $\alpha_{3l}$  & $\alpha_{4l}$ & $\alpha_{2l}$ & $\alpha_{3l}$  & $\alpha_{4l}$    \\ \hline
    4 &  6 &(2.17) &0.664 & 0.770 &(2.17) & 0.664 & 0.770 & (2.17) &0.557& 0.482  \\ 
    4 &  7 &0.890 &0.437 & 0.502 & 0.890 & 0.437 &0.502 &0.890 & 0.376 &0.352   \\
    4 &  8 &0.449 & 0.287&0.319 & 0.448 & 0.287 &0.319 & 0.449&0.255 &0.252   \\ 
    4 &  9 &0.225 & 0.174& 0.184 &0.225  & 0.174 & 0.184 & 0.225 & 0.161&0.162   \\
    4 &  10 & 0.0904&0.0804 &0.0819 & 0.0904 &0.0804  &0.0819 &0.0904  & 0.0775 & 0.0781 \\
   \hline
  \end{tabular}
  \caption{Values of the coupling constant at the infrared fixed point at two, three and four loop level in the $\overline{\text{MS}}$, RI' and mMOM schemes. The fermions are in the two-indexed antisymmetric representation of the $SU(N)$ gauge group.}
\end{center}
\end{table}

\begin{table}[h]
\begin{center}\label{tab:gamma-Antiymmetric}
  \begin{tabular}{|cc||c|c|c||c|c|c||c|c|c|}
  \hline\hline
\multicolumn{2}{|c||}{{\color{rossoCP3}  \rm 2A}}  &   \multicolumn{3}{c||}{$\overline{\text{MS}}$} & \multicolumn{3}{c||}{RI'} & \multicolumn{3}{c|}{mMOM}  \\
\hline
  \hline
    $N$ & $N_f$ & $\gamma_{2l}$ & $\gamma_{3l}$  & $\gamma_{4l}$  & $\gamma_{2l}$  & $\gamma_{3l}$  & $\gamma_{4l}$ & $\gamma_{2l}$ & $\gamma_{3l}$  & $\gamma_{4l}$    \\ \hline
    4 &  6 & (9.78) &1.38  & 0.293 & (13.7) &2.57 &3.03 &(11.3)  &1.57 & 1.81 \\ 
    4 &  7 & 2.19 &0.695 & 0.435 &2.73 &0.942 & 0.565 & 2.48& 0.770 &0.885   \\
    4 &  8 & 0.802 & 0.402 &0.368 &0.904 & 0.449& 0.352 &0.885 & 0.430 &0.477   \\ 
    4 &  9 &0.331 & 0.228&0.232 & 0.348 & 0.234 &0.228  & 0.354 &0.236 &0.248  \\
    4 &  10 & 0.117 & 0.101&0.103 & 0.118 &0.101 &0.102  &0.121 & 0.102 &0.103  \\
    \hline
      \end{tabular}
  \caption{Values of the anomalous dimension of the mass at the infrared fixed point at two, three and four loop level in the $\overline{\text{MS}}$, RI' and mMOM schemes. The fermions are in the two-indexed antisymmetric representation of the $SU(N)$ gauge group.}
\end{center}
\end{table}

\clearpage

\section{Comparison of the Anomalous Dimension at Different Loop Orders}\label{app:ComparisonLoopOrders}

\begin{figure}[h!]
  \centering
     \includegraphics[width=0.25\textwidth]{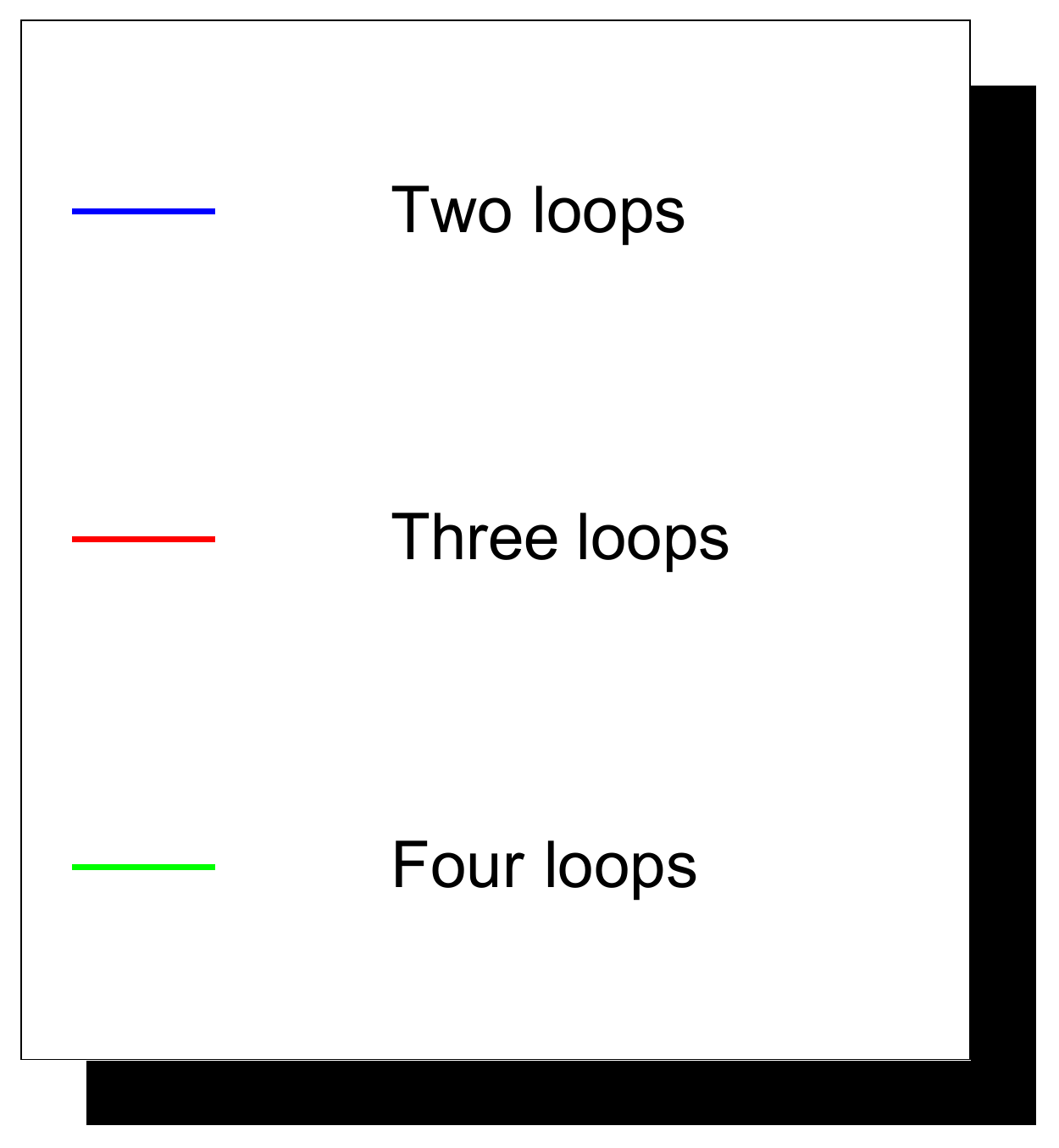}
   \caption{In all the plots below blue curves correspond to two loops, red curves correspond to three loops and green curves correspond to four loops. In addition if the fixed point value $\alpha_0<1$ then the curve is solid while if $\alpha_0>1$ then the curve is dashed.}
\end{figure}

\begin{figure}[h!]
  \centering
    \includegraphics[width=0.4\textwidth]{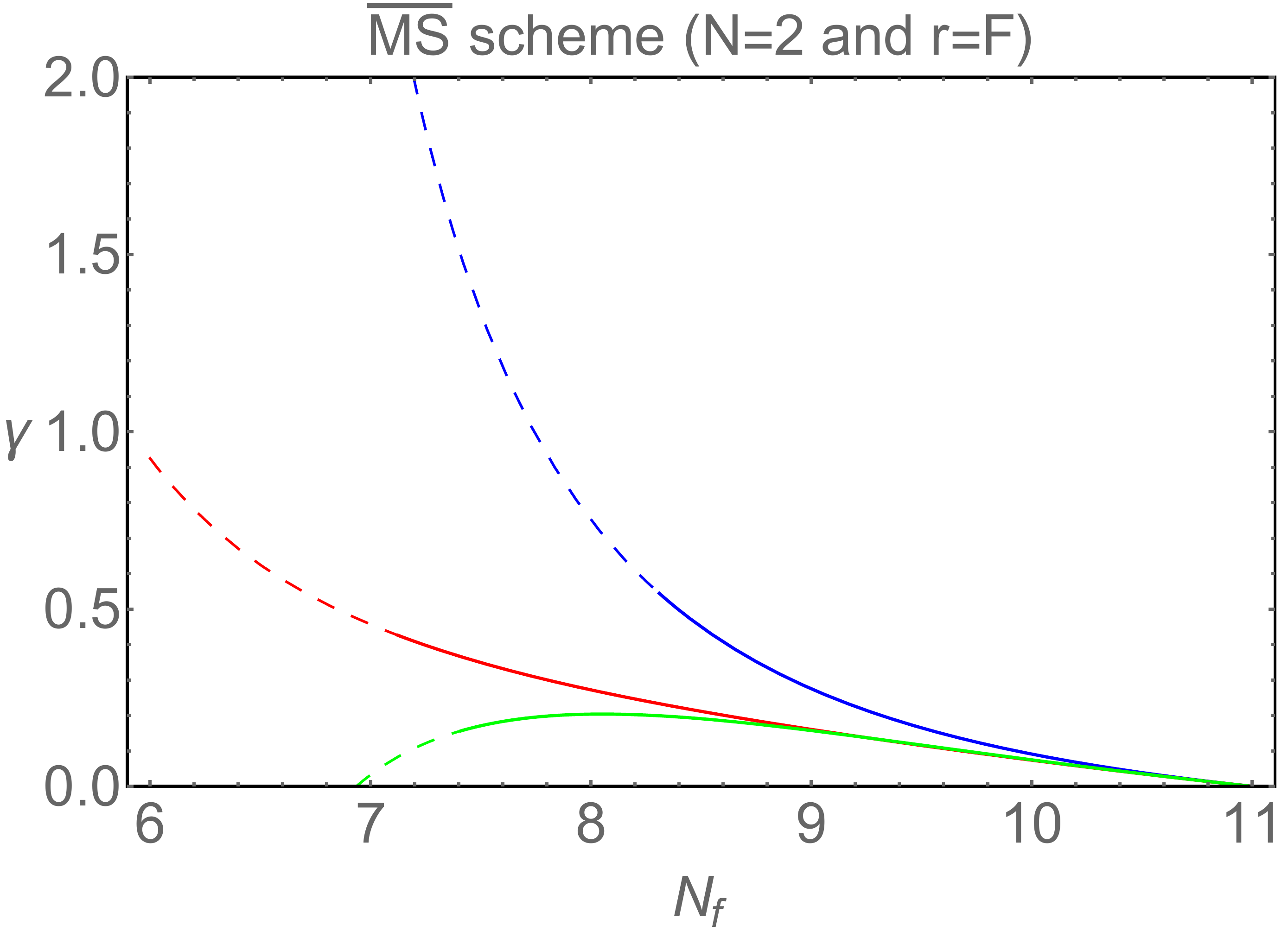}
      \\
     \includegraphics[width=0.4\textwidth]{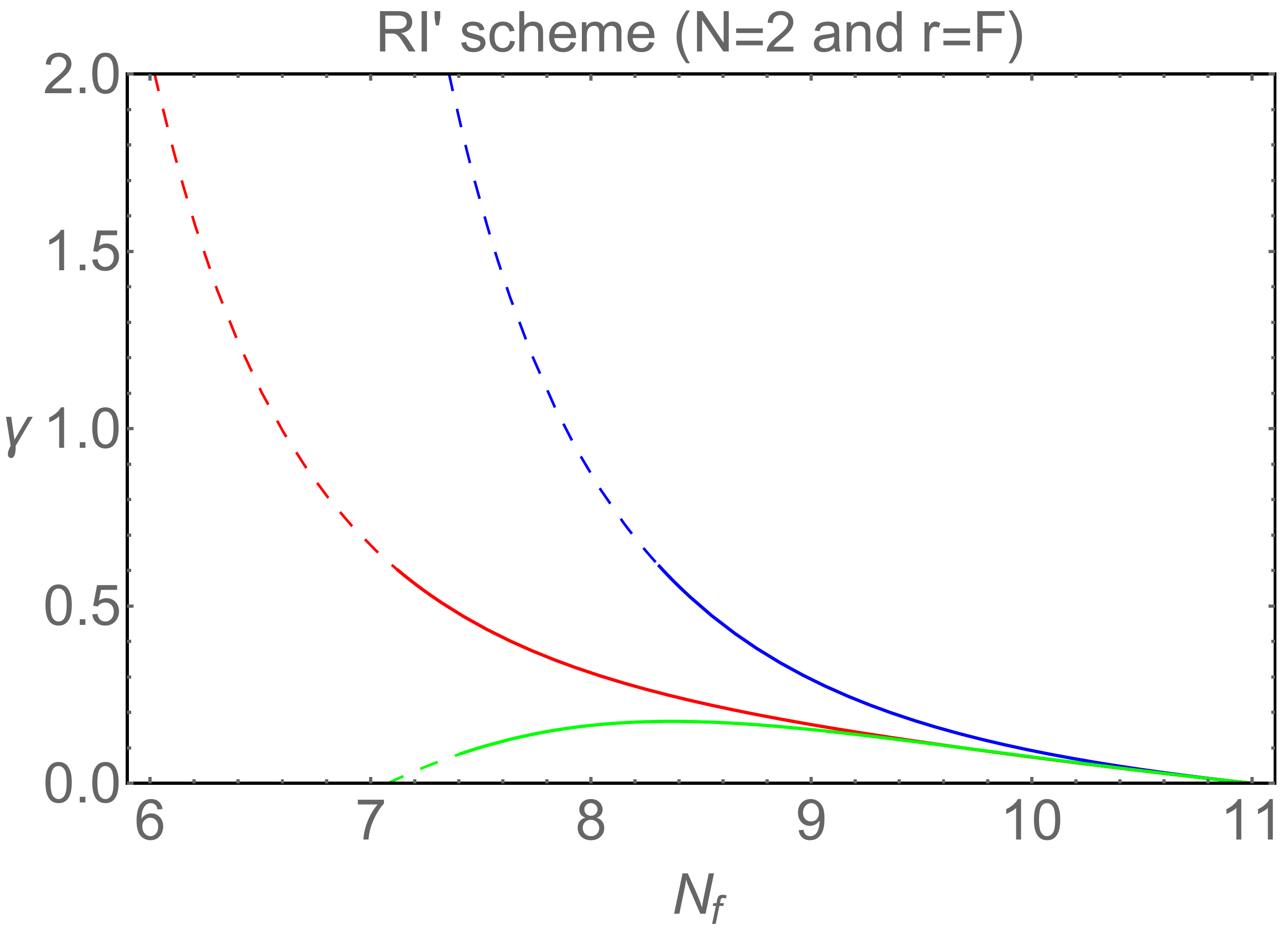} 
     \qquad \qquad
      \includegraphics[width=0.4\textwidth]{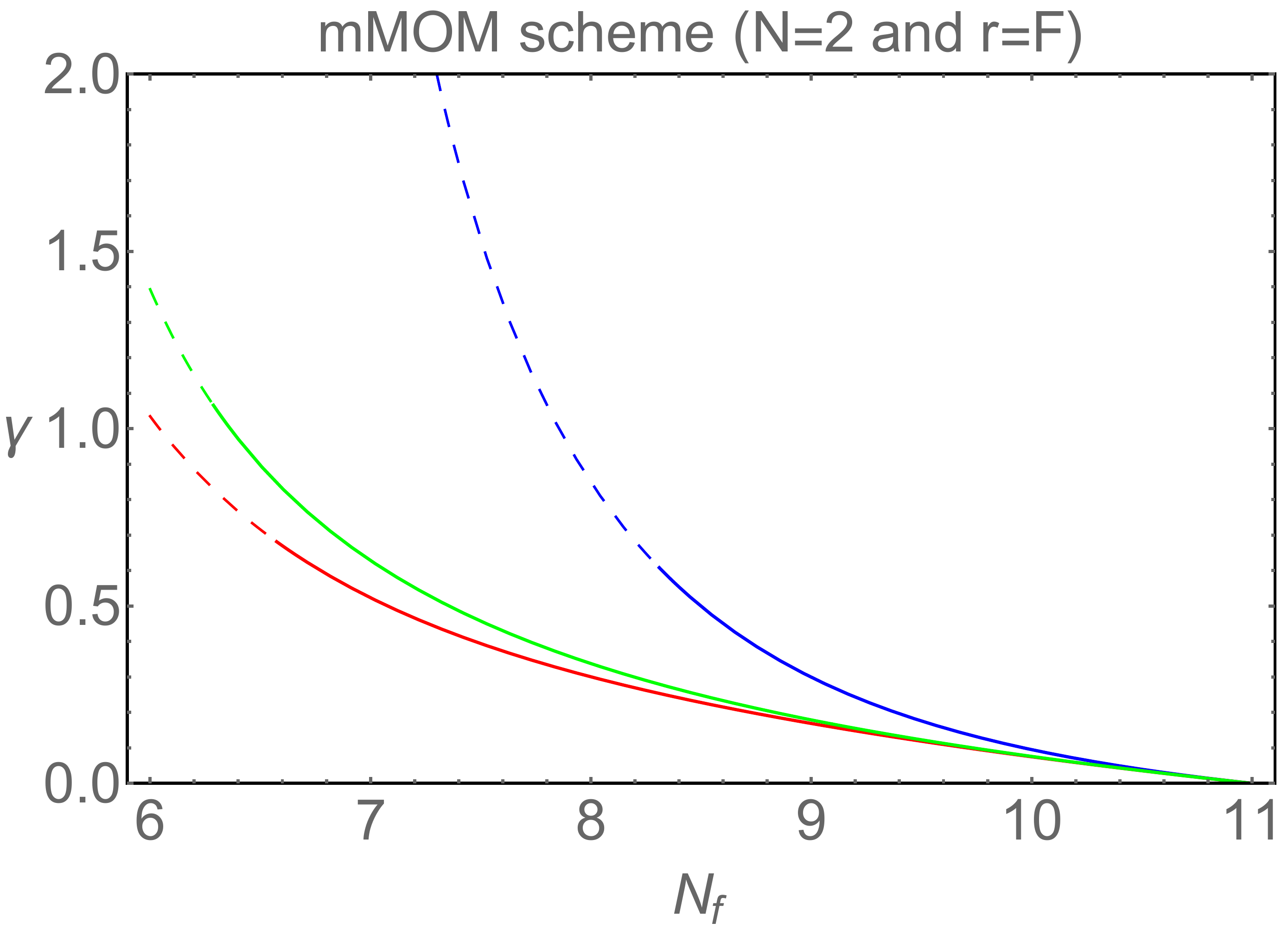}
\caption{The anomalous dimension at an infrared fixed point as a function of the number of flavors. The fermions are in the fundamental representation, F, and the number of colors is $N=2$.}
\end{figure}

\begin{figure}[h!]
  \centering
    \includegraphics[width=0.4\textwidth]{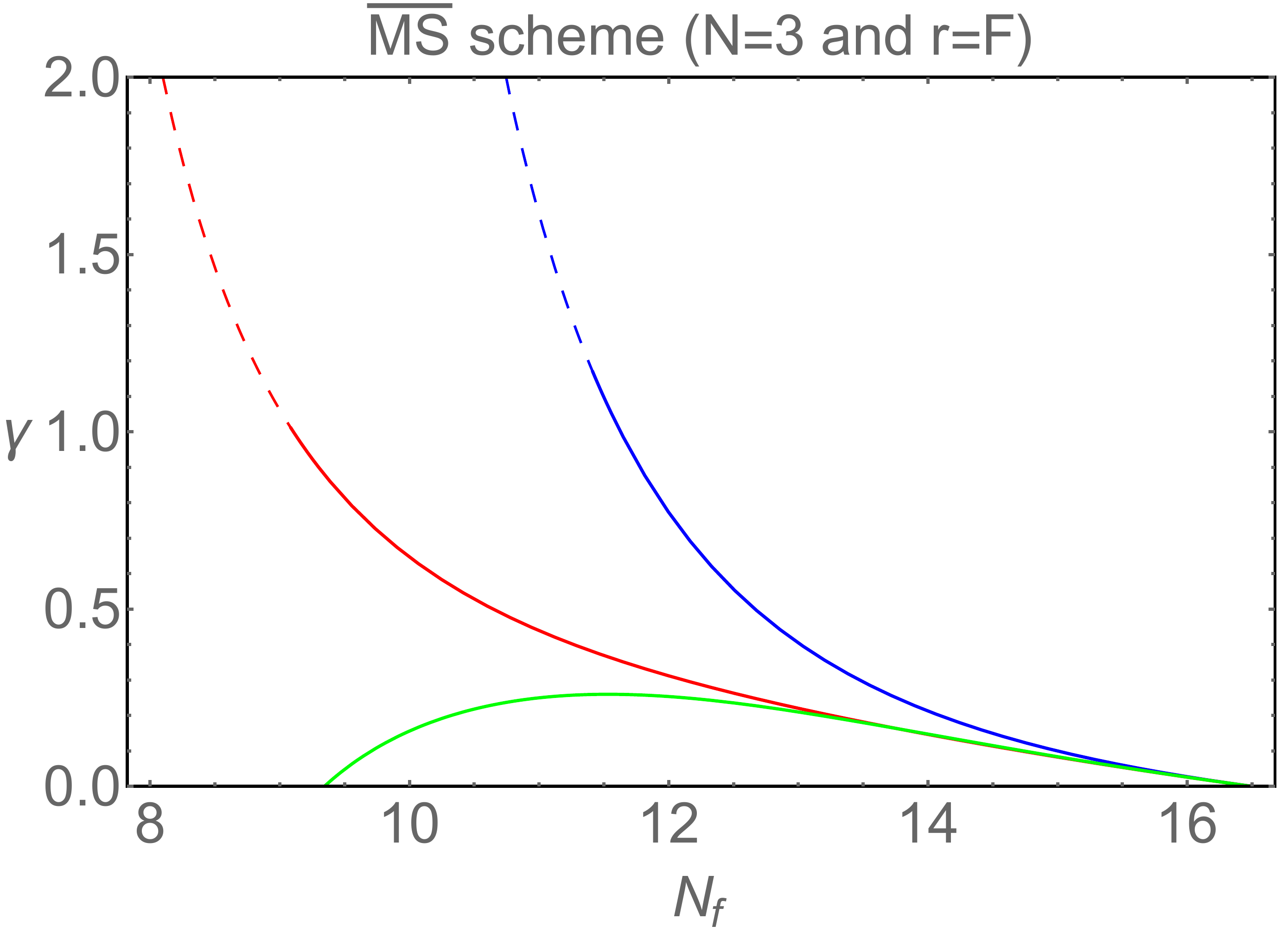}
    \\
     \includegraphics[width=0.4\textwidth]{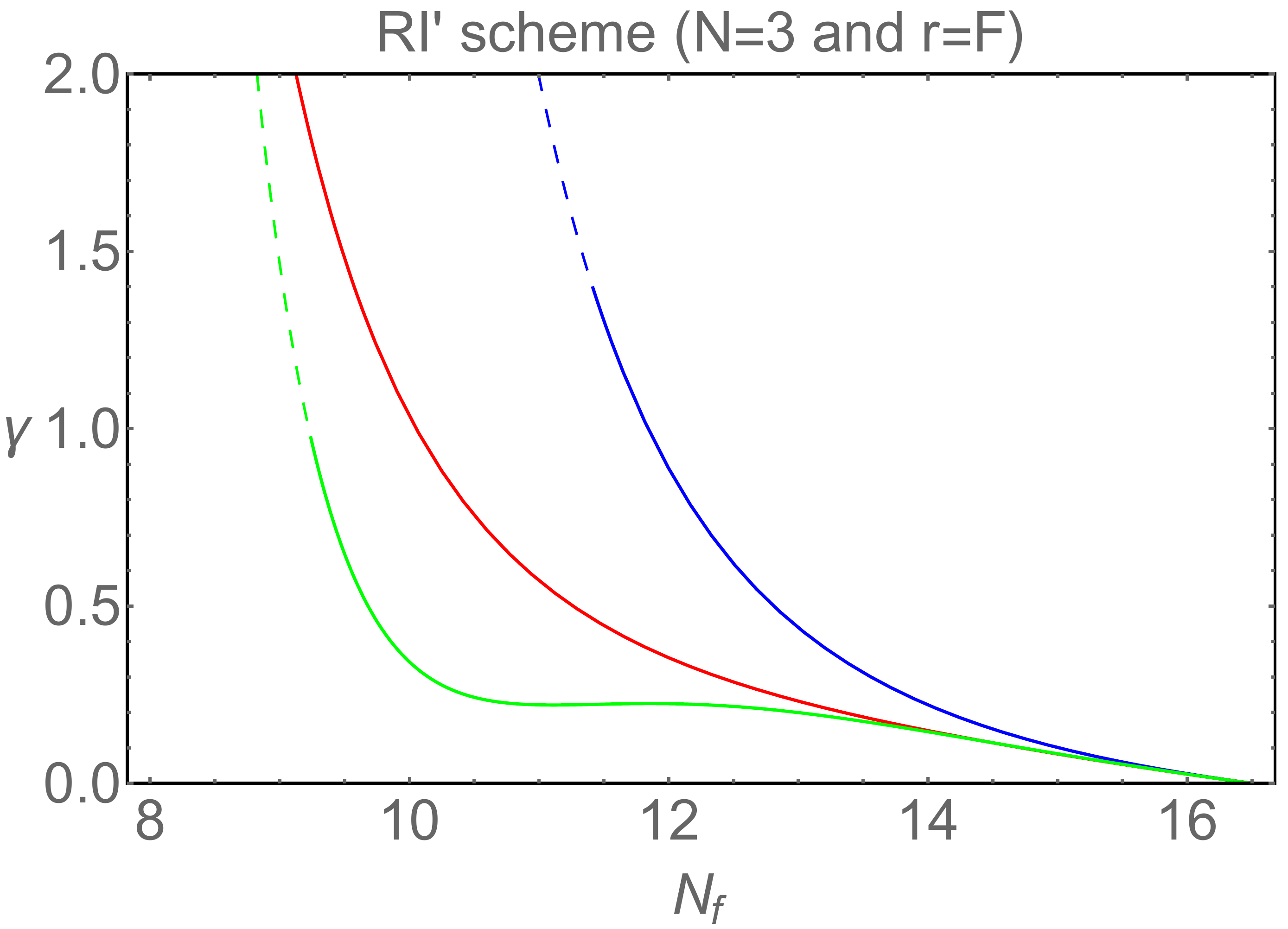} 
     \qquad \qquad
      \includegraphics[width=0.4\textwidth]{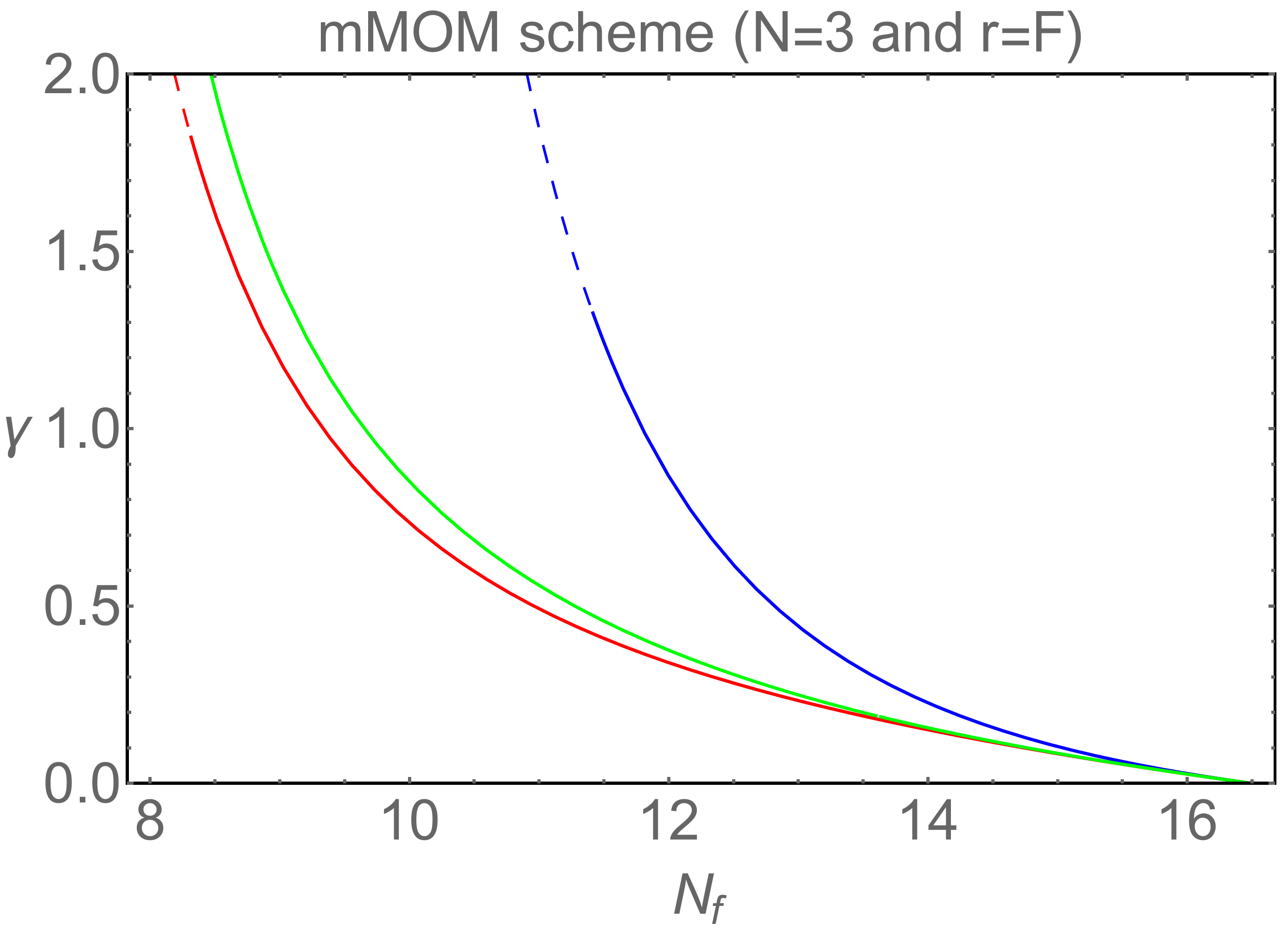}
\caption{The anomalous dimension at an infrared fixed point as a function of the number of flavors. The fermions are in the fundamental representation, F, and the number of colors is $N=3$.}
\end{figure}

\begin{figure}[h!]
  \centering
    \includegraphics[width=0.4\textwidth]{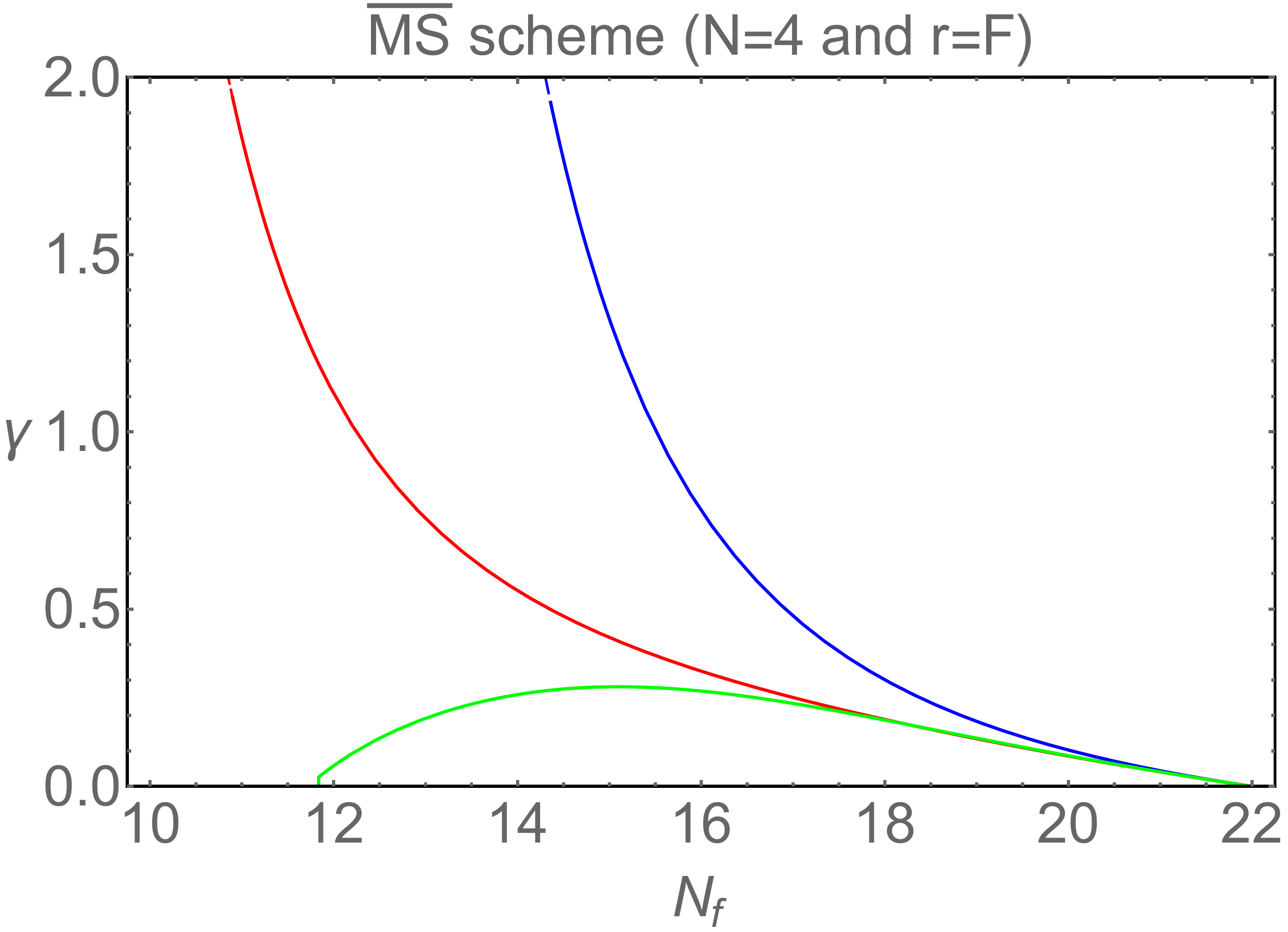}
     \\
     \includegraphics[width=0.4\textwidth]{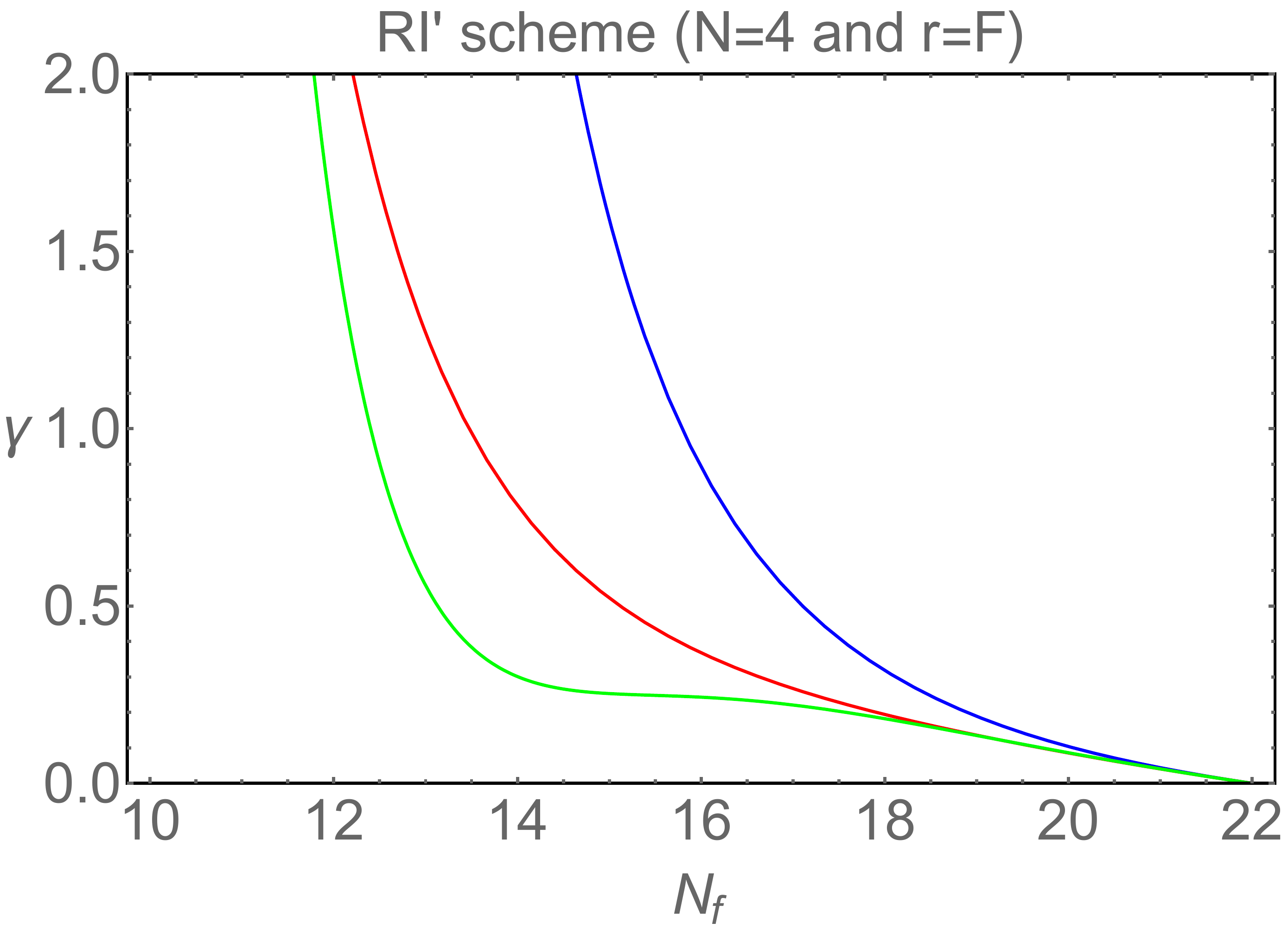} 
     \qquad \qquad
      \includegraphics[width=0.4\textwidth]{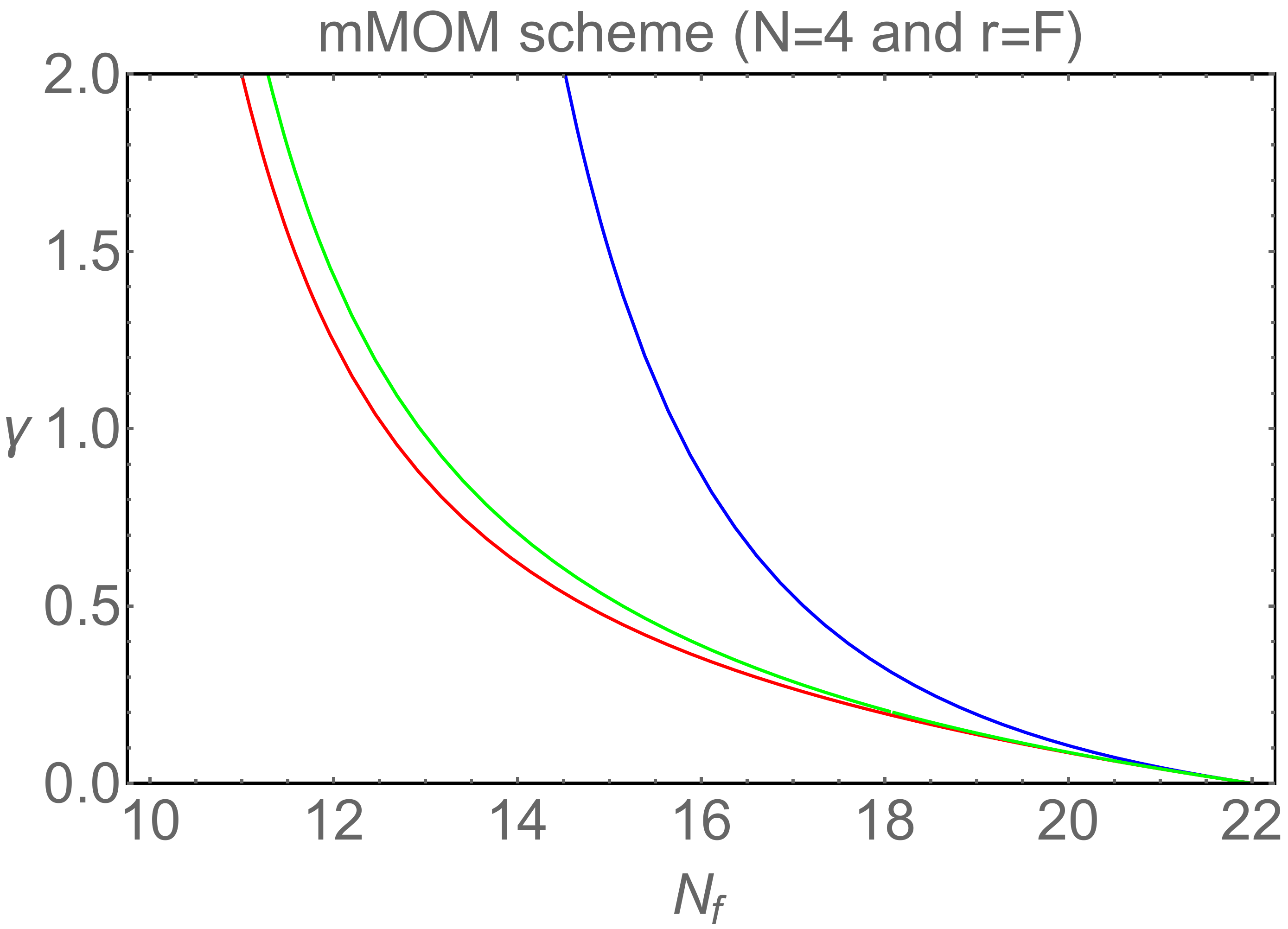}
\caption{The anomalous dimension at an infrared fixed point as a function of the number of flavors. The fermions are in the fundamental representation, F, and the number of colors is $N=4$.}
\end{figure}

\begin{figure}[h!]
  \centering
    \includegraphics[width=0.4\textwidth]{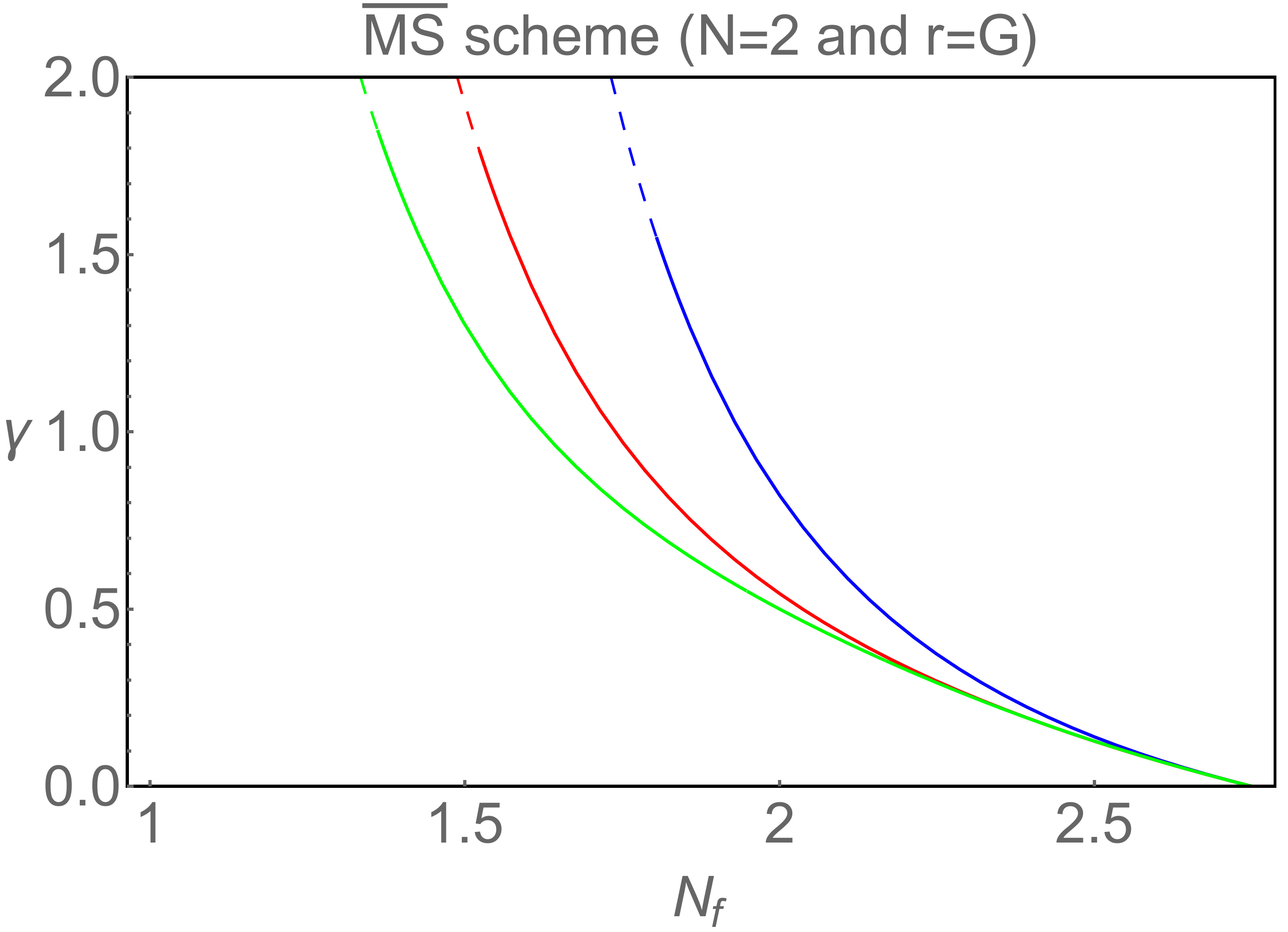}
      \\
     \includegraphics[width=0.4\textwidth]{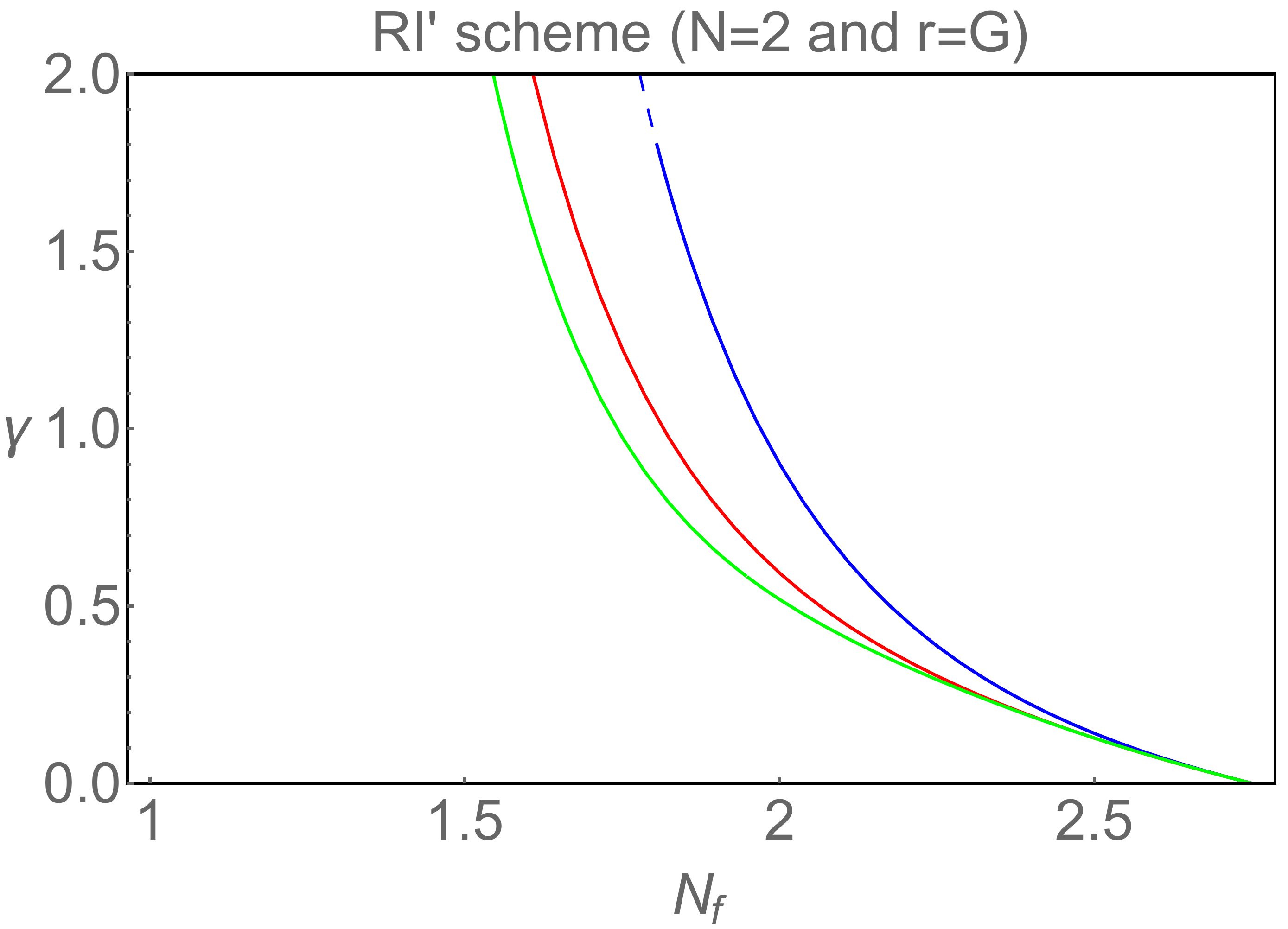} 
     \qquad \qquad
      \includegraphics[width=0.4\textwidth]{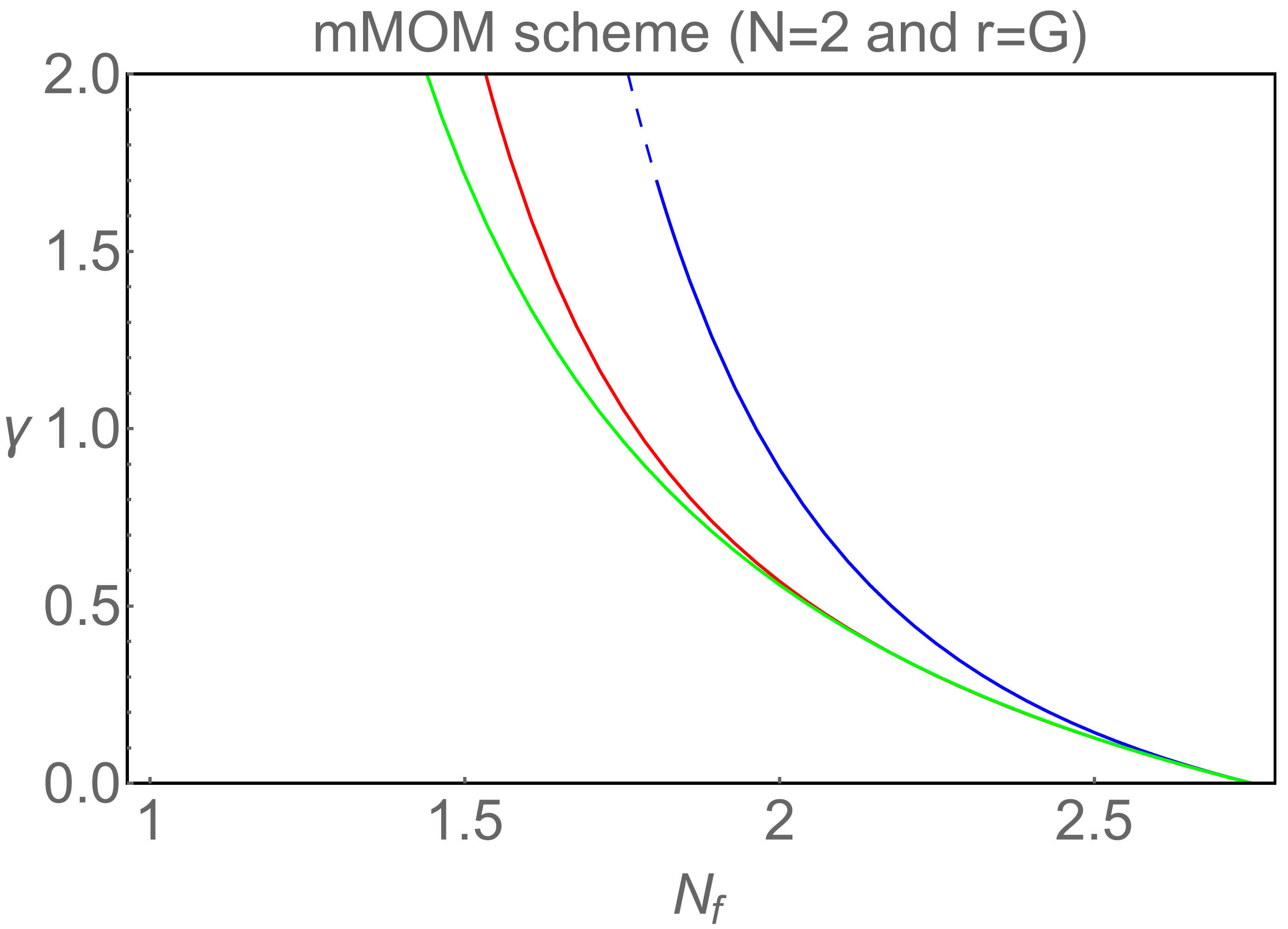}
\caption{The anomalous dimension at an infrared fixed point as a function of the number of flavors. The fermions are in the adjoint representation, G, and the number of colors is $N=2$.}
\end{figure}

\begin{figure}[h!]
  \centering
    \includegraphics[width=0.4\textwidth]{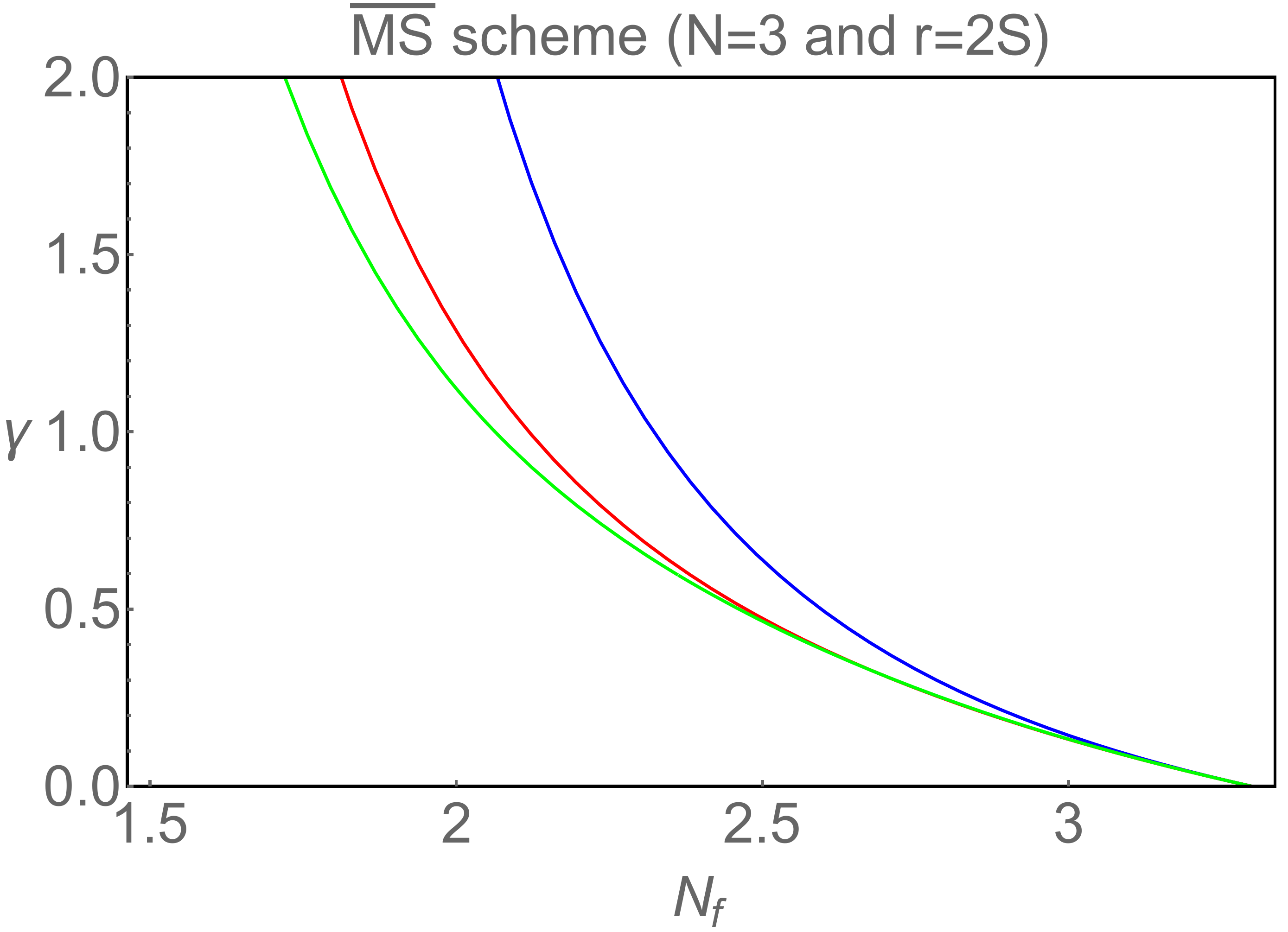}
     \\
     \includegraphics[width=0.4\textwidth]{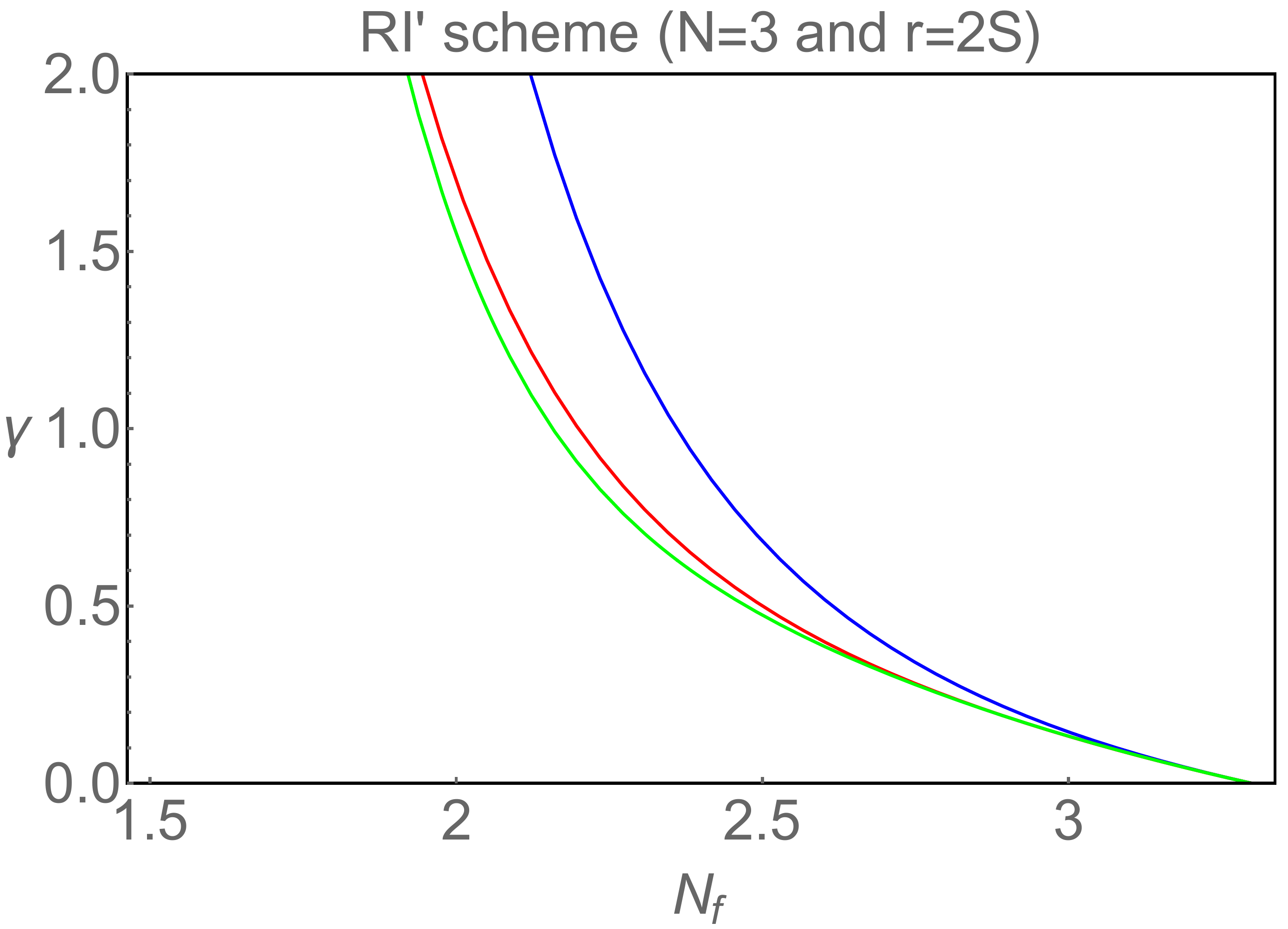} 
     \qquad \qquad
      \includegraphics[width=0.4\textwidth]{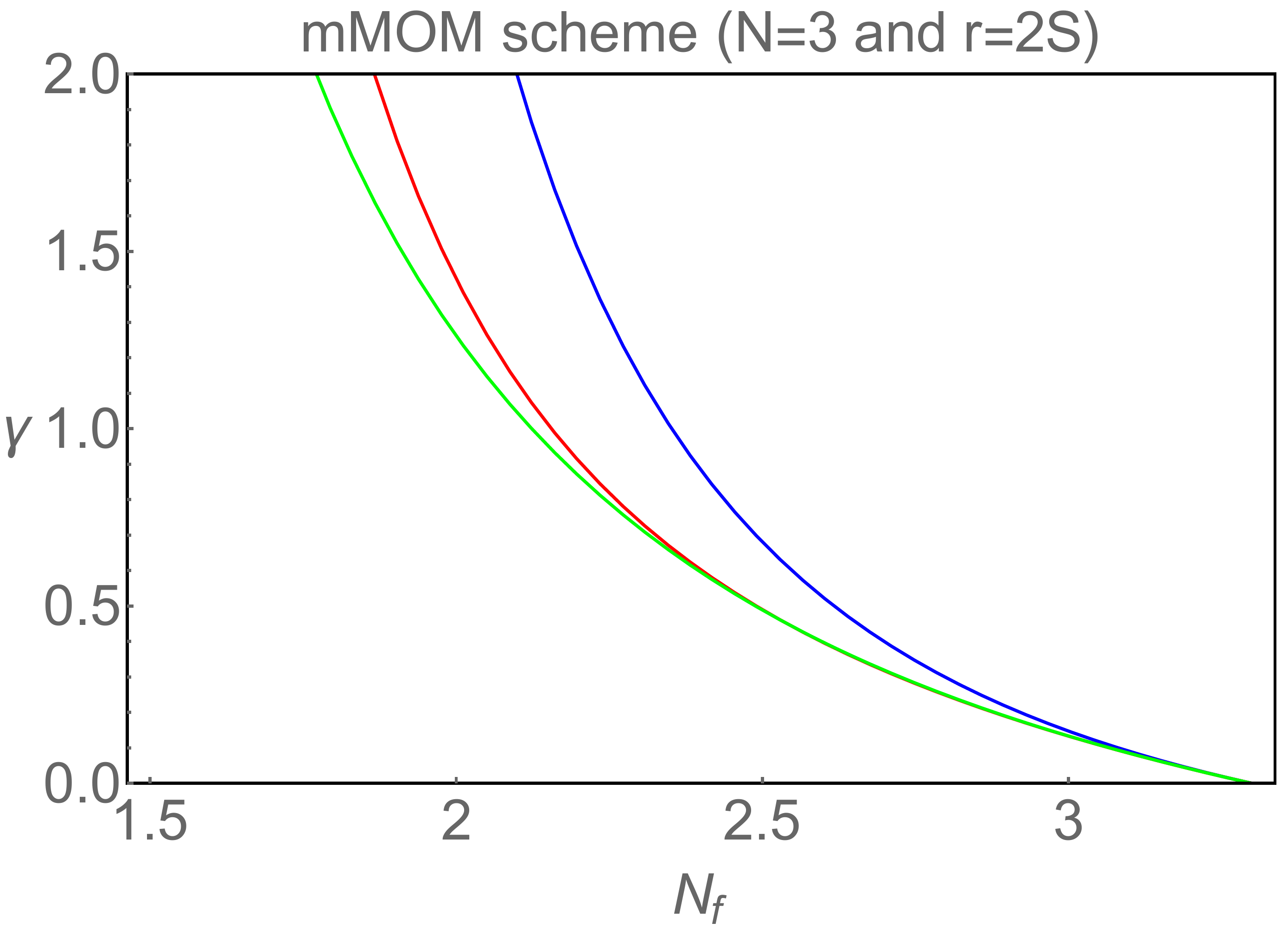}
\caption{The anomalous dimension at an infrared fixed point as a function of the number of flavors. The fermions are in the two-indexed symmetric representation, 2S, and the number of colors is $N=3$.}
\end{figure}

\begin{figure}[h!]
  \centering
    \includegraphics[width=0.4\textwidth]{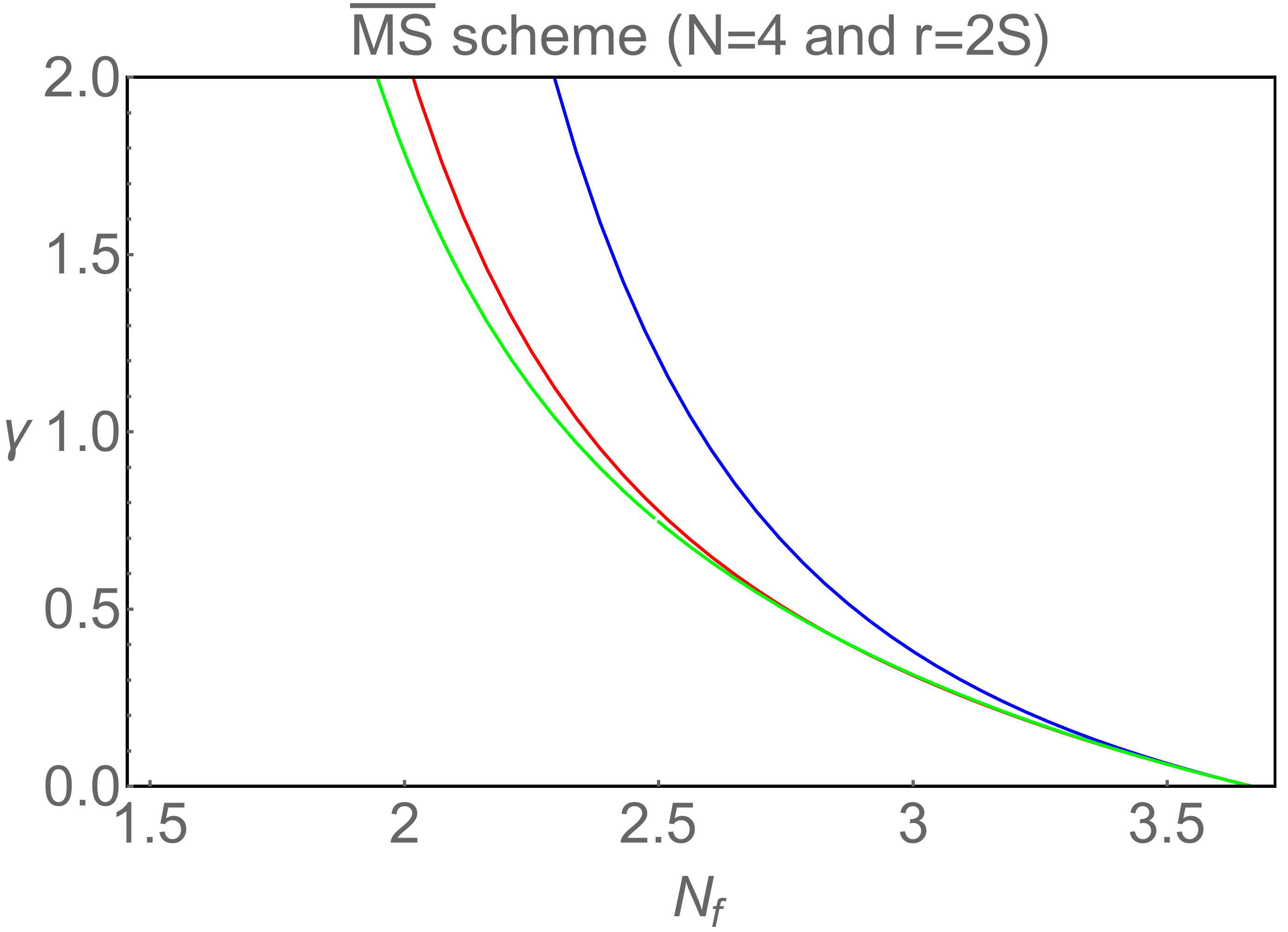}
     \\
     \includegraphics[width=0.4\textwidth]{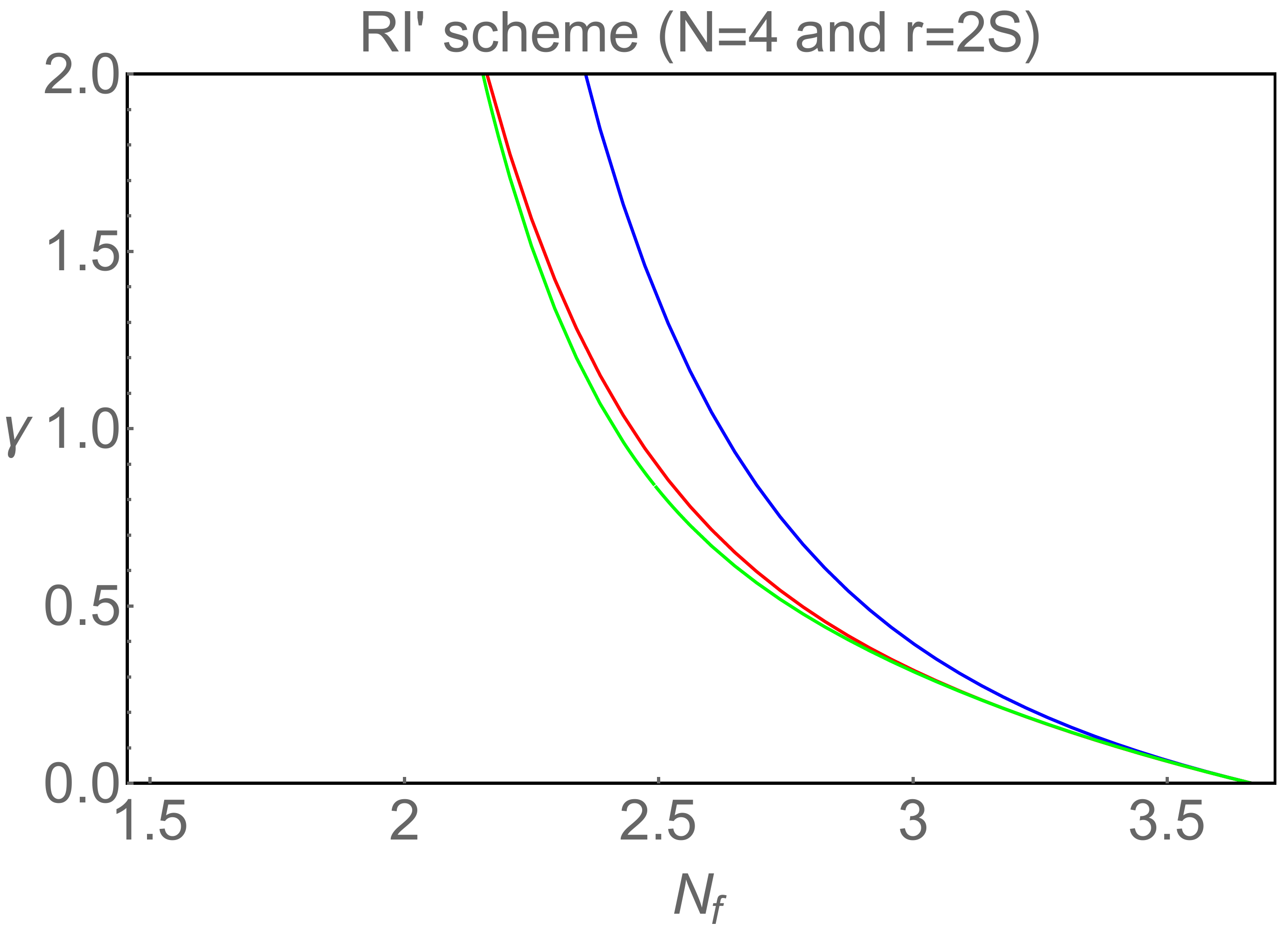} 
     \qquad \qquad
      \includegraphics[width=0.4\textwidth]{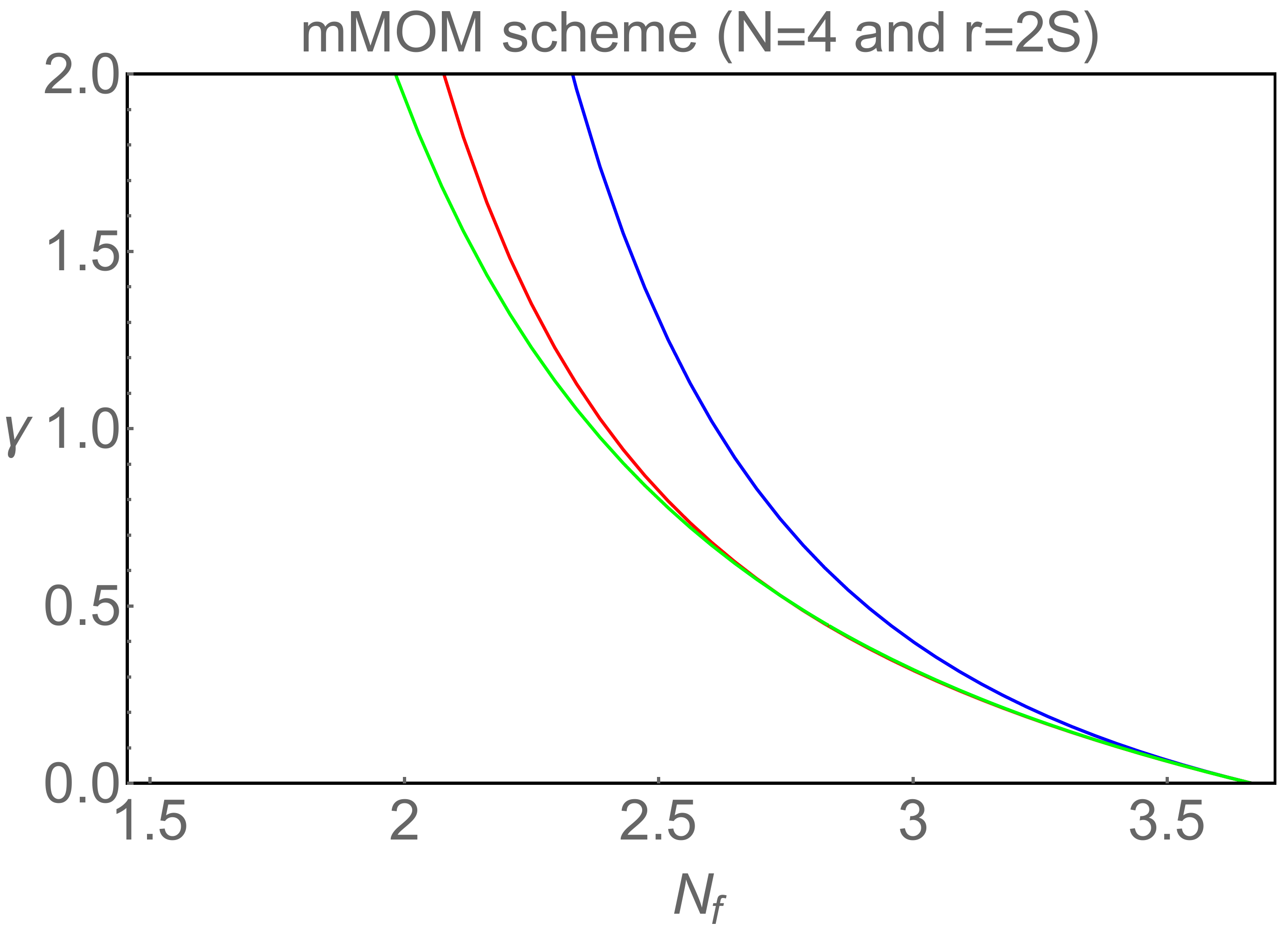}
\caption{The anomalous dimension at an infrared fixed point as a function of the number of flavors. The fermions are in the two-indexed symmetric representation, 2S, and the number of colors is $N=4$.}
\end{figure}

\begin{figure}[h!]
  \centering
    \includegraphics[width=0.4\textwidth]{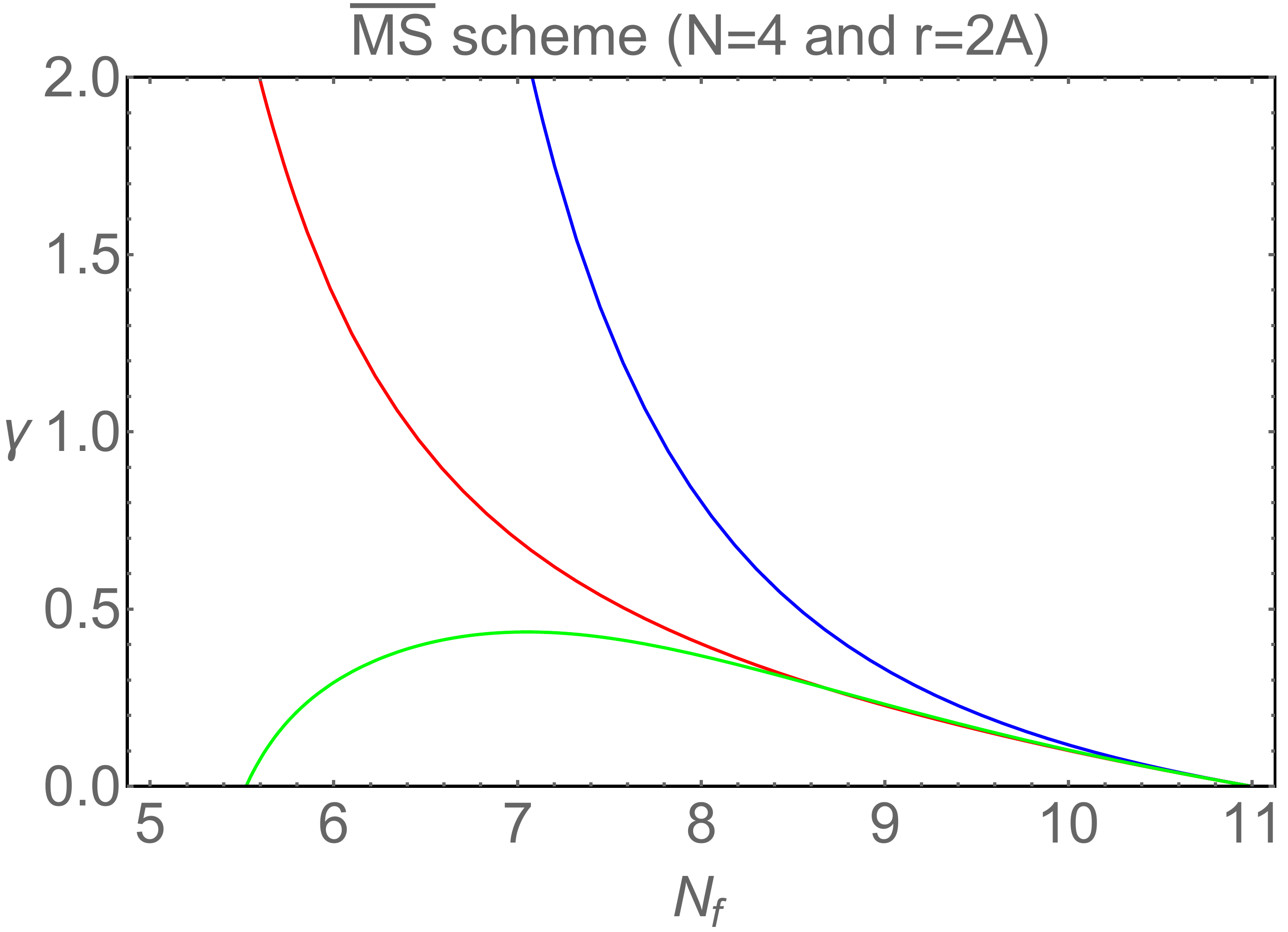}
    \\
     \includegraphics[width=0.4\textwidth]{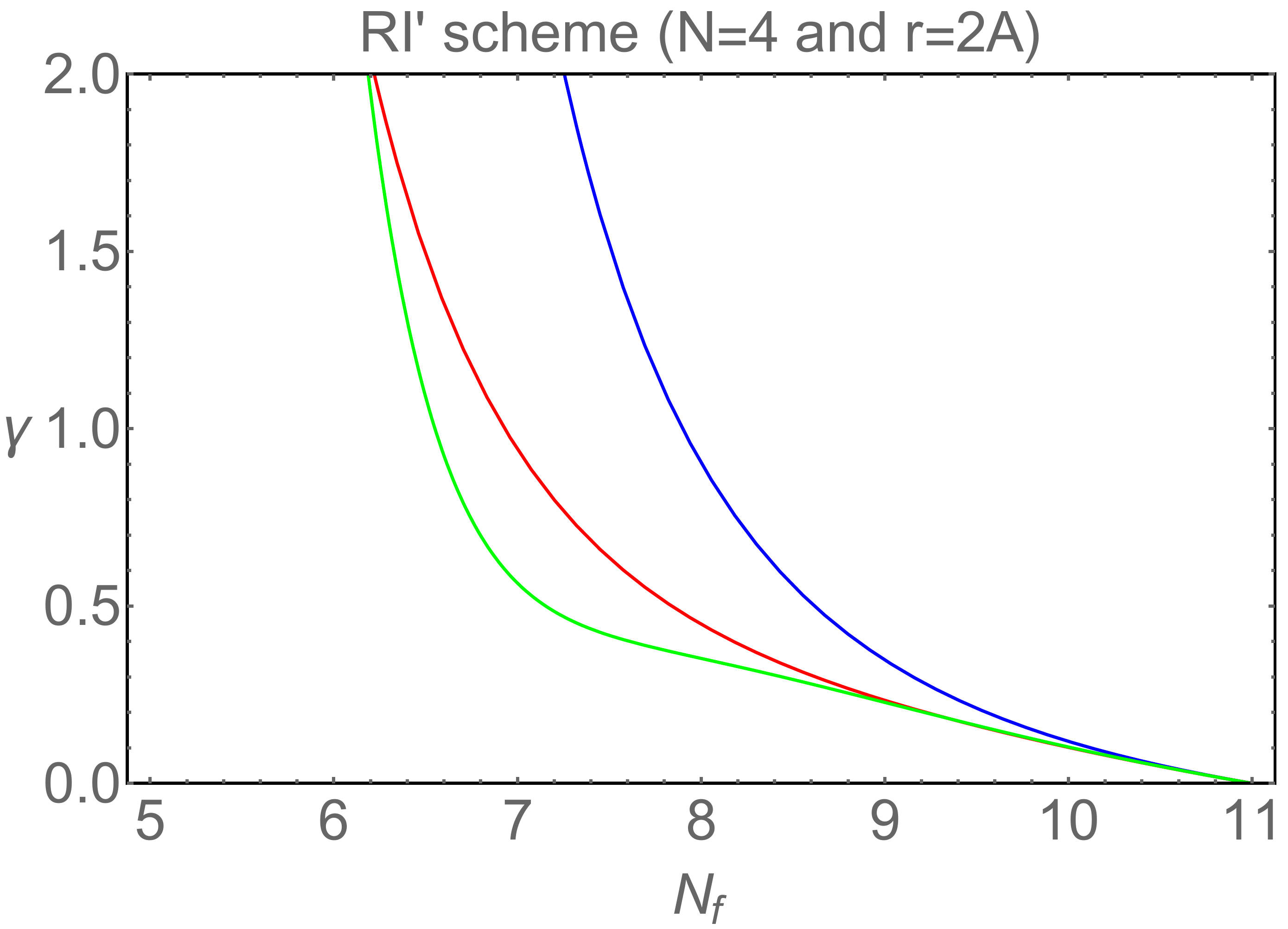} 
     \qquad \qquad
      \includegraphics[width=0.4\textwidth]{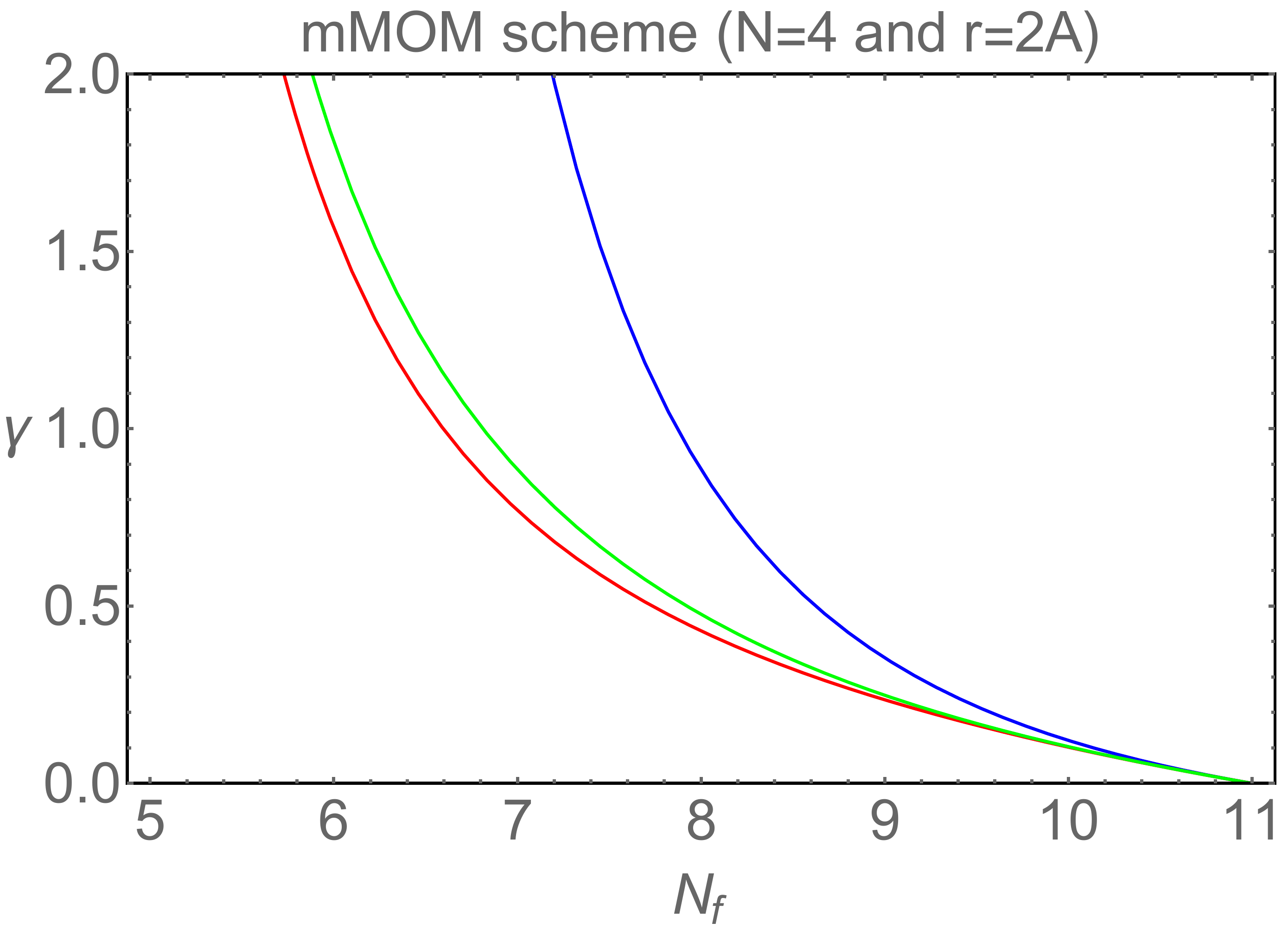}
\caption{The anomalous dimension at an infrared fixed point as a function of the number of flavors. The fermions are in the two-indexed antisymmetric representation, 2A, and the number of colors is $N=4$.}
\end{figure}

\end{document}